\documentclass{aa}  

\usepackage{graphicx}
\usepackage{natbib}
\defcitealias{Pineda2015NatureB5cores}{P15}
\defcitealias{Schmiedeke2021supercriticalfil}{S21}
\usepackage{orcidlink}
\usepackage[normalem]{ulem}
\usepackage{txfonts}
\usepackage[version=3]{mhchem}

\newcommand{\kms}{\ensuremath{\rm km\,s^{-1}}\xspace}
\newcommand{\Msun}{\ensuremath{\rm M_{\sun}}\xspace}

\newcommand{\vlsr}{\ensuremath{\rm \textit{v}_{LSR}}\xspace}
\newcommand{\commentstere}[1]{\textcolor{black}{#1}}
\newcommand{\refereecomment}[1]{\textcolor{black}{#1}}
\newcommand{\refereetwo}[1]{\textcolor{black}{#1}}

\begin{document}

   \title{Flow of gas detected from \refereetwo{beyond the filaments} to protostellar scales in Barnard 5}

   \author{M. T. Valdivia-Mena\inst{1}\orcidlink{0000-0002-0347-3837}
          \and
          J. E. Pineda\inst{1}\orcidlink{0000-0002-3972-1978}
          \and
          D. M. Segura-Cox\inst{2}\thanks{NSF Astronomy and Astrophysics Postdoctoral Fellow}\orcidlink{0000-0003-3172-6763}
          \and
          P. Caselli\inst{1}\orcidlink{0000-0003-1481-7911}
          \and
          A. Schmiedeke\inst{3}\orcidlink{0000-0002-1730-8832}
          \and
          S. Choudhury\inst{4}\orcidlink{0000-0002-7497-2713}
           \and 
          S. S. R. Offner\inst{2}\orcidlink{0000-0003-1252-9916}
          \and
          R. Neri\inst{5}
          \and
          A. Goodman\inst{6}\orcidlink{0000-0003-1312-0477}
          \and 
          G. A. Fuller\inst{7,8}\orcidlink{0000-0001-8509-1818}          
          }

    \institute{Max-Planck-Institut f\"ur extraterrestrische Physik,   Giessenbachstrasse 1, D-85748 Garching, Germany\\
         \email{mvaldivi@mpe.mpg.de}
              \and
              Department of Astronomy, The University of Texas at Austin, 2515 Speedway, Austin, TX 78712, USA
              \and
               Green Bank Observatory, PO Box 2, Green Bank, WV 24944, USA
              \and
              Korea Astronomy and Space Science Institute, 776 Daedeok-daero, Yuseong-gu, Daejeon 34055, Republic of Korea
              \and
              Institut de Radioastronomie Millim\'{e}trique (IRAM), 300 rue de la Piscine, F-38406, Saint-Martin d'H\`{e}res, France
              \and
              Harvard-Smithsonian Center for Astrophysics, 60 Garden St., Cambridge, MA 02138, USA
              \and
              Jodrell Bank Centre for Astrophysics, Department of Physics and Astronomy, School of Natural Sciences, The University of Manchester, Oxford Road, Manchester M13 9PL, United Kingdom
              \and
              I. Physikalisches Institut, Universität zu Köln, Zülpicher Str.77, 50937 Köln, Germany
              }

   \date{Received 9 March 2023, Accepted 19 July 2023}

 
  \abstract
  {The infall of gas from outside natal cores has proven to feed protostars after the main accretion phase (Class 0). This changes our view of star formation to a picture that includes asymmetric accretion (streamers), and a larger role of the environment. However, the connection between streamers and the filaments that prevail in star-forming regions is unknown.} 
   {We investigate the flow of material toward the filaments within Barnard 5 (B5) and the infall from the envelope to the protostellar disk of the embedded protostar B5-IRS1. Our goal is to follow the flow of material from the larger, dense core scale, to the protostellar disk scale. }
   {We present new \ce{HC3N} line data from the NOEMA and 30m telescopes covering the coherence zone of B5, together with ALMA \ce{H2CO} and \ce{C^{18}O} maps toward the protostellar envelope. We fit multiple Gaussian components to the lines so as to decompose their individual physical components. We investigated the \ce{HC3N} velocity gradients to determine the direction of chemically fresh gas flow. At envelope scales, we used a clustering algorithm to disentangle the different kinematic components within \ce{H2CO} emission. }
   {At dense core scales, \ce{HC3N} traces the infall from the B5 region toward the filaments. \ce{HC3N} velocity gradients are consistent with accretion toward the filament spines plus flow along them. We found a $\sim 2800$ au streamer in \ce{H2CO} emission, which is blueshifted with respect to the protostar and deposits gas at outer disk scales. The strongest velocity gradients at large scales curve toward the position of the streamer at small scales, suggesting a connection between both flows.}
   {Our analysis suggests that the gas can flow from the dense core to the protostar. This implies that the mass available for a protostar is not limited to its envelope, and it can receive chemically unprocessed gas after the main accretion phase.}


   \maketitle
%

\section{Introduction}

Stars form in dense cores, which are defined as local overdensities at a sub-parsec scale with respect to their background. These contain cold ($\sim 10$ K) and dense ($n\geq10^{5}$ cm$^{-3}$) gas from which an individual star or a bound stellar system \commentstere{could} form \citep{Bergin-Tafalla2007,Pineda2022PPVII}. \commentstere{The classical picture of low-mass star formation ($\lesssim 1$ \Msun) considers a core as an isolated unit with initial angular momentum that collapses to form a protostar and a protostellar disk \citep{Shu1977, Terebey1984rotation}.} However, cores do not live in isolation: observations of star-forming regions show that they are mostly harbored within filaments, which continuously accrete gas from the larger molecular cloud and evolve \citep{Andre2010Herschel,Konyves2015Herschel-coresandfils}. Filaments themselves are highly \commentstere{structured} and present preferential velocity-coherent \commentstere{flows of gas within them}, called fibers \citep[e.g.,][]{Hacar2013TaurusFibers, Hacar2017NGC1333Fibers}. Observations show that material tends to collapse to the densest parts of filaments \commentstere{on} $\sim 0.1$ pc scales ($\sim 20\,000$ au) and toward cores while star formation is underway \citep{Andre2010Herschel,Arzoumanian2011Herschel-fil-width}. Therefore, the assumption of isolation and symmetry breaks when looking at cores harbored within filaments, and understanding the kinematics of dense cores allows us to trace the origin of gas that forms a protostar. In particular, infall phenomena are vital to understand the growth and final masses of protostars and their planetary systems.

Recently, it has been found that at smaller scales ($\sim1000$ au) within the core, infall also proceeds in an asymmetric manner, through channels called streamers \citep{Pineda2020Per2streamer, Garufi2021arXiv-accretionDGTauHLTau,Valdivia-Mena2022per50stream}. Streamers are found from the highly embedded Class 0 phase \citep{LeGouellec2019PolarizedClass0,Pineda2020Per2streamer} 
to Class II sources \citep[e.g.,][]{ Yen2019HLTau, Akiyama2019,Alves2020doublestream,Garufi2021arXiv-accretionDGTauHLTau,Ginski2021}. They have also been found feeding not only single protostars, but also protostellar binaries, both funneling material toward the inner circumstellar disks \citep{Phuong2020GGTauSpirals,Alves2019PretzelAccretion, Dutrey2014GGTau} and to the binary system as a whole \citep{Pineda2020Per2streamer}. It has been suggested that streamers can deliver material from beyond the dense core \citep{Kuffmeier2023rejuvenatinginfall}, bringing even more material toward protostars \commentstere{than} that supplied from their own envelopes \citep{Pineda2020Per2streamer,Valdivia-Mena2022per50stream}. These results show that star formation is an asymmetric and chaotic process driven by the motions of gas within a molecular cloud. A new \commentstere{picture of} low-mass star formation is thus emerging, where the relationship of the protostar and the local environment, together with asymmetries, \commentstere{is} much more relevant \citep{Pineda2022PPVII}.

\commentstere{While} our understanding of both streamers and infall toward filaments is growing, \commentstere{if and how these infall mechanisms connect at large and small scales is} still an open question. For this work, we probed two scales of infall, first the infall of fresh gas toward the filaments and condensations, and then the infall of gas from a protostellar envelope to a protostar. For this, we studied the dense core region of Barnard 5 (B5), regarded as a quiescent \commentstere{core at a distance of 302 pc from our Solar System \citep{Zucker2018distance}}, located at the eastern edge of the Perseus molecular cloud. 
Barnard 5 has been the subject of several studies, most of them tracing the dense gas structure using  \ce{NH3}, but also through other tracers such as CO, \ce{C^{18}O}, \ce{N2H+}, and HCN \citep{Fuller1991B5anatomy,Pineda2010coherenceB5, Pineda2015NatureB5cores,Schmiedeke2021supercriticalfil}. \commentstere{We refer to \cite{Pineda2015NatureB5cores} as \citetalias{Pineda2015NatureB5cores} and \cite{Schmiedeke2021supercriticalfil} as \citetalias{Schmiedeke2021supercriticalfil}}. Within this core, there is a clear coherence region, defined as the area where the nonthermal velocity dispersion of dense gas is subsonic \citep{Pineda2010coherenceB5}. Inside this coherent zone, there are two filaments seen in \ce{NH3}, \commentstere{each} about 0.3 pc ($\sim60\, 000$ au) long (Fig.\,\ref{fig:HC3N-integrated}). These filaments have supercritical masses per unit length, indicating that they are not supported against gravitational collapse and that they are currently fragmenting  \citepalias{Schmiedeke2021supercriticalfil}. Within these filaments, there are three condensations likely to form stars and a Class I protostar, B5-IRS1, which together will form a wide-separation (more than 1000 au) quadruple system \citepalias{Pineda2015NatureB5cores}. 

B5-IRS1, also known as Per-emb-53, is identified as a Class I protostar from its spectral energy distribution \citep[SED,][]{Enoch2009yso-properties}. It is located at the northern edge of Fil-2, between Cond-2 and 3 \commentstere{\citepalias[using the nomenclature from][]{Pineda2015NatureB5cores, Schmiedeke2021supercriticalfil}} and it has a central velocity $\vlsr=10.2$ \kms \citepalias{Pineda2015NatureB5cores}. It has a disk, which \commentstere{remains spatially un}resolved, with an estimated mass of \refereecomment{} $0.03$ \Msun\ at most, \refereetwo{using the mass found by} \cite{Zapata2014B5outflow} and correcting for a distance of 302 pc to B5 \citep{Zucker2018distance}. Its outflow cone is almost perpendicular to the orientation of the filament it is embedded in  \refereecomment{\citep[Fig. \ref{fig:HC3N-integrated},][]{Zapata2014B5outflow}}.

This paper is organized as follows. In Sect. \ref{sec:observations}, we describe the data from several different telescopes we used for this work and how they were \commentstere{processed}. In Sect. \ref{sec:results} we describe the new data cubes \commentstere{and how we discovered individual velocity components in the spectra.} Section \ref{sec:analysis} explains how we analyzed the \commentstere{discovered velocity components} and determined the kinematic properties of \commentstere{B5 and the envelope surrounding IRS1}. In Sect. \ref{sec:discussion} we discuss our results and connect the (large-scale) \commentstere{kinematic properties of the filaments within B5} to the (small-scale) \commentstere{infalling gas in the \refereecomment{protostellar} envelope}. Section \ref{sec:summary} summarizes \commentstere{the} main results and conclusions of our work.


\section{Observations and data reduction\label{sec:observations}}

\begin{table*}[!htbp]
        \centering
        \caption{\label{tab:obsprops}Properties of the molecular line observations \refereecomment{and telescopes} used in this work.}
        \begin{tabular}{ccccccc}
                \hline\hline   
                Molecule & Transition             & Frequency$^1$  & Telescope   & Spatial resolution              & Spectral resolution  & rms  \\
                &              &  (GHz) & & (\arcsec)             &  (km s$^{-1}$) &  (K) \\
                \hline
                \ce{HC3N}  & 8 -- 7                & 72.783822    & NOEMA + 30m & $5.13\times4.70$  & \commentstere{0.257}                              & 0.17    \\
                \ce{HC3N}  & 10 -- 9                & 90.979023       & NOEMA + 30m & $4.07\times3.79$  & \commentstere{0.206}                              & 0.15    \\
                \ce{H2CO}  & 3$_{0,3}$ -- 2$_{0,2}$ & 218.222195      & ALMA        & $0.41 \times 0.27 $ & \commentstere{0.084}                              & 0.98    \\
                \ce{C^{18}O} & 2 -- 1                 & 219.560354      & ALMA        & $0.40 \times 0.27 $ & \commentstere{0.042}                              &    1.47     \\
                
                \hline
        \end{tabular}
        \tablefoot{The rms is a mean value. \tablefoottext{1}{\commentstere{Frequencies taken from the  Cologne Database for Molecular Spectroscopy \citep[CDMS,][]{Endres2016CDMS}.}}} 
\end{table*}

We used observations from different telescopes to \commentstere{investigate the kinematics} of \commentstere{the two} B5 filaments within the coherent core, and the envelope around B5-IRS1 \citep[3\textsuperscript{h}47\textsuperscript{m}41.591\textsuperscript{s}, +32\degr51\arcmin43.672\arcsec (J2000), ][]{Tobin2016VANDAMmultiples}. We used \ce{HC3N}\,($10-9$) and ($8-7$) line observations taken on the Northern Extended Millimeter Array (NOEMA) at Plateau the Bure (France) and the 30m telescope at Pico Veleta (Spain), from Institut de Radioastronomie Milim\'etrique (IRAM). We also used \ce{H2CO}\,($3_{0,3}- 2_{0,2}$) and \ce{C^{18}O}\,($2 - 1$) line cubes observed with the Atacama Large Millimeter/Submillimeter Array (ALMA) at the Chajnantor Plateau (Chile). We refer to \ce{H2CO}\,($3_{0,3}- 2_{0,2}$) and \ce{C^{18}O}\,($2 - 1$) as \ce{H2CO} and \ce{C^{18}O} for the rest of this work, respectively. A summary of the molecular transitions used in this work and the data properties are in Table \ref{tab:obsprops}. \commentstere{Additionally, we used \ce{NH3}\,(1,1) line observations and spectral fit from \citetalias{Pineda2015NatureB5cores}, taken with the Karl G. Jansky Very Large Array (VLA) in New Mexico and with the Green Bank Telescope (GBT) in West Virginia (USA). Details about these observations and fits can be found in Appendix \ref{sec:obs-vla+gbt}.}

\subsection{NOEMA\label{sec:obs-noema} }

NOEMA observations were carried out under project S18AL (PI: J. E. Pineda) using the Band 1 receiver. B5 was observed on 2018 August 24 and between 2018 September 15 and September 23 in D configuration, using a mosaic with 53 pointings. \commentstere{We used the PolyFix correlator tuned with a LO frequency of 82.499 GHz. The \ce{HC3N}\,($8-7$) and ($10-9$) line frequencies (Table \ref{tab:obsprops}) are located at high resolution chunks, with a channel width of 62.5 kHz.} For the observing period between August 24 and September 16, the source was observed with eight antennas, while the rest of the period there were ten antennas available. The data were calibrated using the standard observatory pipeline in the Continuum and Line Interferometer Calibration (CLIC) program, which is part of the Grenoble Image and Line Data Analysis Software (GILDAS) package.

\subsection{30m Telescope\label{sec:obs-30m}}

Observations of Barnard 5 with the IRAM 30 telescope were \commentstere{obtained} with the EMIR 090 receiver. The observations were made under project 034-19 (PIs: J. E. Pineda and A. Schmiedeke), between 2019 August 24 and 26, using on-the-fly mapping and connected to the FTS50 backend. The 30m data were reduced using \commentstere{CLASS}. We used \refereetwo{J0319+4130} for pointing \commentstere{and the data are calibrated using the two-load method \citep{Carter2012EMIRtwoload}.}

\subsection{Combination of NOEMA and 30m\label{sec:obs-combo}}

We used the GILDAS software \texttt{mapping} to combine and image the NOEMA and 30m data of \commentstere{both \ce{HC3N} molecular transitions used in this work}. \commentstere{The command \texttt{uv\_short} is used to combine the 30-m and NOEMA data into a combined uv-table.} We used natural weight and the multiscale CLEAN algorithm implemented in \texttt{mapping} to obtain a CLEANed datacube. We manually did the mask using the \texttt{support} command in \texttt{mapping}. The final properties of the \ce{HC3N} data cube\commentstere{s are} in Table \ref{tab:obsprops}. Finally, we integrated the obtained cube\commentstere{s} to make velocity integrated image\commentstere{s} from 9.2 to 11.2 \kms, so as to cover the full range of velocities where emission has a signal-to-noise ratio $(S/N)>3$.

\subsection{ALMA\label{sec:obs-alma}}

We observed the envelope surrounding B5-IRS1 at high resolution with ALMA \commentstere{Band 6} under project 2017.1.01078.S (PI: D. Segura-Cox). Observations were done during 2018 September 22 with the 12m array using \commentstere{49 antennas}. The total integration time was 20.16 minutes. The phasecenter of our observations is 3\textsuperscript{h}47\textsuperscript{m}41.588\textsuperscript{s}, +32\degr51\arcmin43.643\arcsec (J2000). \commentstere{The minimum baseline length was 15.07 m, resulting in a maximum recoverable scale (MRS) of 4\arcsec, and maximum baseline of 1398 m.} \commentstere{The primary beam for our observations is approximately 15\farcs4 for both molecules. }

We used the Common Astronomy Software Applications  \citep[CASA,][]{McMullin2007CASA} version 5.4.0-68 \commentstere{for data calibration. We used the calibration results from the pipeline ran by the ALMA OSF.} We used \commentstere{J0237+2848  and J0510+1800 as bandpass calibrators. J0336+3218 was used by the pipeline for gain calibration. The data presented here are self-calibrated in an iterative process, with phase-only self-calibration having a shortest solution interval of the integration time (6.05s) and a final round including amplitude self-calibration with an infinite solution interval.}

We imaged the data using CASA version 6.4.0. The \ce{H2CO} and \ce{C^{18}O} spectral cubes are produced using robust (Briggs) weight with a parameter $r=0.5$ to balance flux sensitivity and resolution. We used the \texttt{tclean} procedure through a manually selected mask on the visible signal. We first use the classic Hogbom CLEAN algorithm to create the masks. For the final cubes, we used multiscale CLEAN with scales [0, 5, 25, 50] pixels (which correspond to scales of [0, 0.27, 1.35, 2.7]\arcsec~ in angular units, approximately [0, 0.7, 3 and 7] times the beam) for the \ce{H2CO} emission, and the same for \ce{C^{18}O} plus 100 pixels, which corresponds to 5\farcs2. This significantly reduces the presence of artifacts (or regions of negative emission) due to the presence of extended emission that we cannot fully recover. The final properties of the \ce{H2CO} and \ce{C^{18}O} emission cubes are in Table \ref{tab:obsprops}. 

After imaging, we found that the calibration process over-subtracted continuum emission around the position of the protostar. This effect is small and reduces the baseline of line emission only by 0.5 K on average. To correct for this, we took the median value of each spectrum in channels that do not have emission and subtract this median value to each spectrum. \commentstere{As the median value around the protostar's position is negative, this subtraction raises the values of the spectra. }

\section[Results]{Results\label{sec:results}\footnote{All codes used to obtain the following results can be accessed through \href{https://github.com/tere-valdivia/Barnard_5_infall}{Github}.}}

\subsection{\commentstere{Morphology of \ce{HC3N} emission}}

\begin{figure}[htbp]
        \centering
        \includegraphics[width=0.49\textwidth]{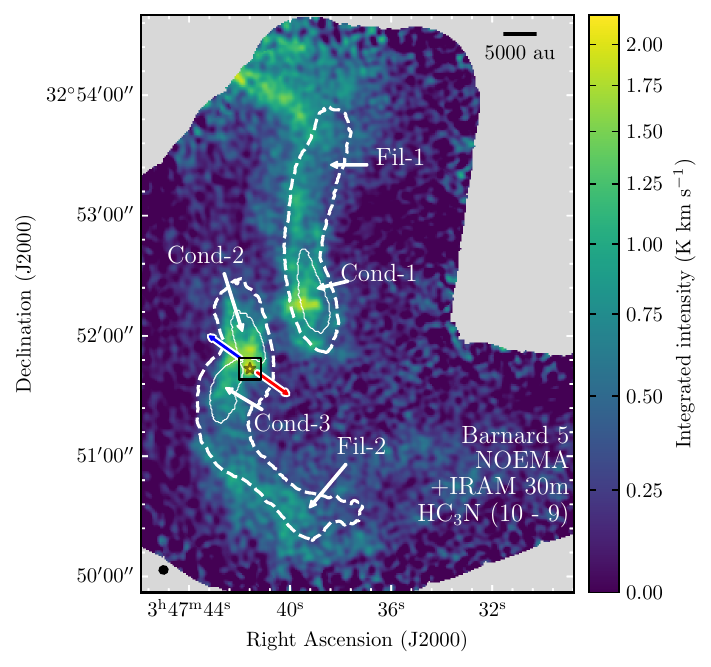}
        \caption{Velocity integrated \ce{HC3N} $(10-9)$ emission. \ce{HC3N} is integrated from 9.2 to 11.2 \kms. White dashed contours correspond to the filaments identified in \ce{NH3} emission by \citetalias{Pineda2015NatureB5cores} and \citetalias{Schmiedeke2021supercriticalfil}. White \commentstere{solid} contours \commentstere{outline} the edges of the condensations labeled as in \citetalias{Pineda2015NatureB5cores}. The \refereecomment{gray} \commentstere{star} marks the position of the protostar B5-IRS1. The black \refereecomment{square} represents the area \commentstere{observed} with ALMA. The blue and red arrows indicate the direction of the blueshifted and redshifted lobes of the outflow  \refereetwo{in B5-IRS1 \citep{Zapata2014B5outflow}}.}
        \label{fig:HC3N-integrated}
\end{figure}

\begin{figure*}[htbp]
        \centering
        \includegraphics[width=0.95\textwidth]{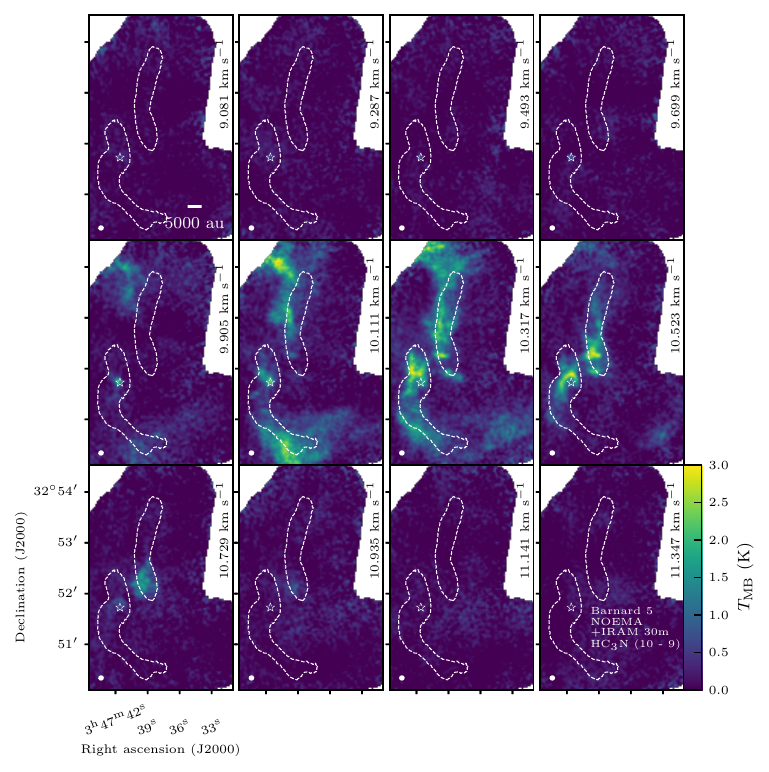}
        \caption{Channel maps between 9.081 and 11.347 \kms for the NOEMA and 30m \ce{HC3N} (10 -- 9) spectral cube, in 0.206 \kms steps. The white dashed contours \commentstere{outline the \ce{NH3}} filaments defined in \citetalias{Pineda2015NatureB5cores}. The white star represents the location of the protostar, \refereecomment{which has a systemic velocity $\vlsr=10.2$ \kms \citepalias{Pineda2015NatureB5cores}}.  The white ellipse at the bottom left corner represents the size of the beam. \commentstere{The white line in the top left channel map is a scalebar representing a length of 5000 au.} }
        \label{fig:HC3N-chanmap}
\end{figure*}

Figure \ref{fig:HC3N-integrated} shows the integrated \commentstere{intensity} of \ce{HC3N}\,($10-9$) emission. We overplot the contours from the \ce{NH3} condensations and filaments from \citetalias{Pineda2015NatureB5cores} and \citetalias{Schmiedeke2021supercriticalfil}, \refereecomment{ data which we further describe in Appendix~\ref{sec:obs-vla+gbt}.} \ce{HC3N}\,($8-7$) emission shows the same morphology, so this description applies to the ($8-7$) transition as well.  \commentstere{The \ce{HC3N}\,($8-7$) integrated intensity map is shown in Appendix \ref{sec:HC3N87line}.}  In general, \ce{HC3N}\,($10-9$) follows the shape of the \ce{NH3}\,(1,1) filaments, but its emission is more extended than the filaments, with a width of \commentstere{$\sim 40\arcsec$}. Toward the northeast, the \ce{HC3N}\,($10-9$) emission continues beyond the edge of Fil-1 with a bright peak ($S/N>10$) near the edge of the NOEMA map, possibly connecting to the region defined as Clump 1 in \citetalias{Schmiedeke2021supercriticalfil}. \commentstere{We} note that, with the exception of the extended emission toward the north,  \ce{HC3N} emission is not detected in the extended coherent core defined in \cite{Pineda2010coherenceB5}. The peak of \ce{HC3N}\,($10-9$) emission is located toward B5-IRS1, with two dimmer peaks in integrated emission, one centered \refereetwo{on} Cond-1 and the other at the southern edge of Cond-2. 

Figure \ref{fig:HC3N-chanmap} shows the \ce{HC3N}\,($10-9$) channel maps in \commentstere{approximately} 0.2 \kms steps. These channel maps show that \ce{HC3N}\,($10-9$) emission consists of two lanes of extended emission, one lane for each filament seen in \ce{NH3}\,(1,1), and peaks of emission located at the condensations, the protostar and toward the north of Fil-1 in the direction of Clump-1. Emission along the filaments is usually detected within 3 to 4 channels, except at locations such as \commentstere{Cond-1 and} the protostar, where emission is \commentstere{present} in more than 5 channels.

\subsection{\commentstere{Morphology of \ce{H2CO} line emission}}

\begin{figure}[htbp]
        \centering
        \includegraphics[width=0.49\textwidth]{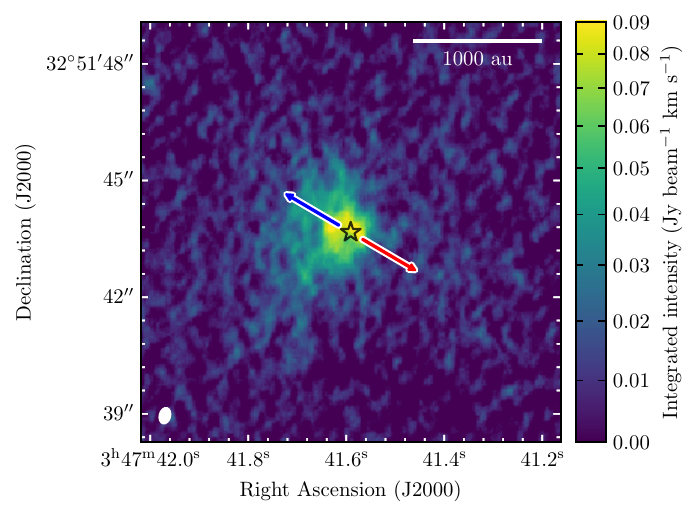}
        \caption{Velocity integrated image of \ce{H2CO} from 8 to 12 \kms. The black star represents the position of the protostar. The blue and red arrows indicate the directions of the blue and redshifted outflow lobes from \cite{Zapata2014B5outflow}. The scalebar indicates a length of 1000 au. The white ellipse in the bottom left corner represents the beam size. The primary beam of the image is \refereetwo{out of the map.}}
        \label{fig:H2CO-integrated}
\end{figure}

\begin{figure*}[hbtp]
        \centering
        \includegraphics[width=0.99\textwidth]{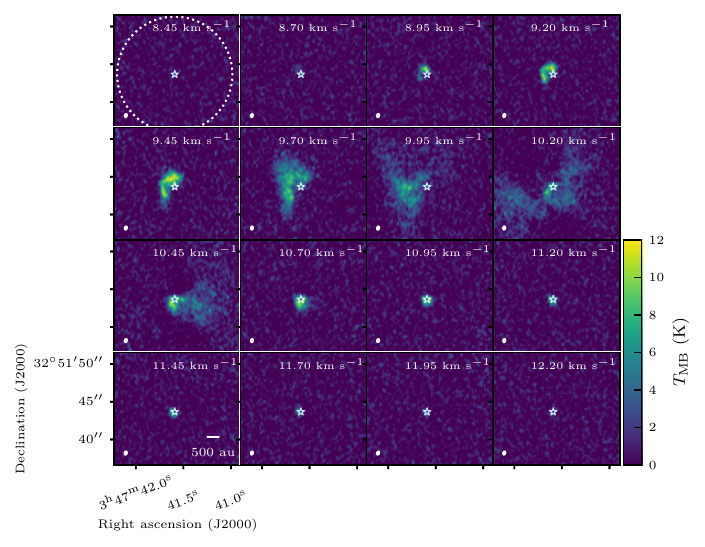}
        \caption{Channel maps between 8.45 and 12.2 \kms for the ALMA \ce{H2CO}\,(3$_{0,3}$ -- 2$_{0,2}$) spectral cube, \commentstere{in steps of 0.25 \kms}. The white star represents the location of the protostar, \refereecomment{which has a systemic velocity $\vlsr=10.2$ \kms \citepalias{Pineda2015NatureB5cores}}. The white dotted circle marks the extent of the primary beam. The white ellipse at the bottom left corners represents the beam size. The scalebar \refereetwo{in the bottom left panel} shows a 500 au length in the map.}
        \label{fig:H2CO-chanmap}
\end{figure*}

Figure \ref{fig:H2CO-integrated} shows the \ce{H2CO} integrated image from 8 to 12 \kms, and Fig. \ref{fig:H2CO-chanmap} shows channel maps of \ce{H2CO} emission in \commentstere{approximately} 0.2 \kms steps. \refereecomment{The details of the observed \ce{C^{18}O} emission morphology are described in Appendix \ref{ap:C18Odata}.} The integrated image peaks toward the position of B5-IRS1 and shows stronger emission toward the east of the protostar than to the west, immediately hinting at asymmetries in the envelope. In the channel maps it is shown that the asymmetry is due to strong emission surrounding the protostar (within a $\sim500$ au diameter) only toward the northeast, showing a curved shaped emission between \refereecomment{9.20} and 9.54 \kms, which then decreases in intensity but expands toward the east in consecutive channels as it reaches the central velocity of the protostar $\vlsr=10.2$ \kms \citepalias{Pineda2015NatureB5cores}. Channels from 9.7 to 10.3 \kms are affected by negative emission bowls adjacent to the extended emission, due to missing short-spacing data in the uv-plane. \commentstere{In the rest of the channels emission is more compact, and it is well recovered with the ALMA only observations.} In higher velocity channels, \commentstere{from 10.45 to 10.95 \kms \refereecomment{in fig. \ref{fig:H2CO-chanmap}},} there is asymmetric envelope emission toward the west of the protostar that is not evident in the integrated image. \refereecomment{These \refereetwo{features} are evident in the image moments 1 and 2, which are shown in Appendix \ref{ap:H2COmoments}. The peak of velocity dispersion is offset from the protostar to the east, where two sections with different velocities are joined. Also, there is an increase in velocity dispersion toward the southeast of the protostar, where emission is primarilly blueshifted. The moment maps and channel maps reveal that \ce{H2CO} emission shows complex structure both spatially and spectrally. }

\subsection{\ce{HC3N} (10 -- 9) line fitting \label{sec:HC3Nfit}}

\begin{figure*}[htbp]
        \centering
        \includegraphics[width=0.99\textwidth]{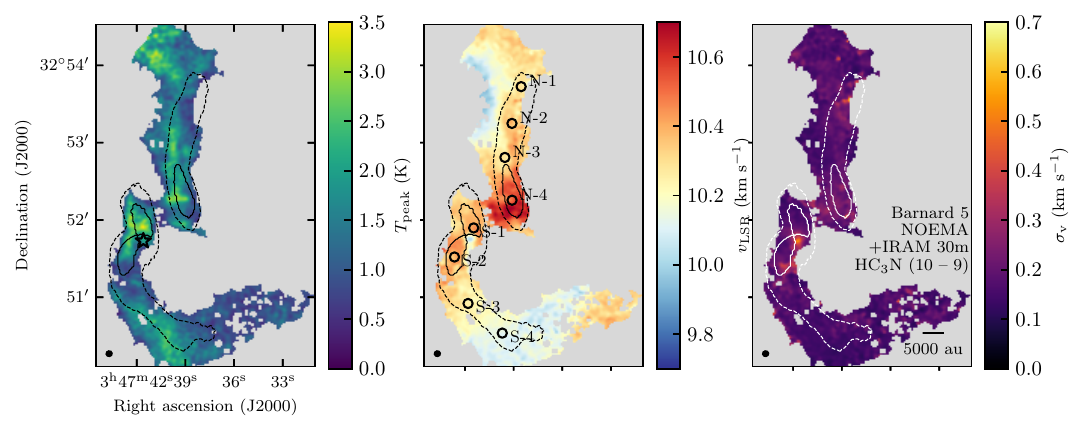}
        \caption{Peak intensity, central velocity, and velocity dispersion resulting from the Gaussian fit to the \ce{HC3N}\,($10-9$) spectra. Black and white dashed contours represent the edges of the filaments as defined in \citetalias{Pineda2015NatureB5cores} and \citetalias{Schmiedeke2021supercriticalfil}. \commentstere{The black and white solid contours mark the} condensations as defined in \citetalias{Pineda2015NatureB5cores}. The black star marks the position of B5-IRS1. \refereecomment{Black} labeled circles indicate the position of the sampled spectra in Fig. \ref{fig:HC3Nspec}. The black ellipse represents the beam size. }
        \label{fig:HC3N-fitparams}
\end{figure*}

\begin{figure}
        \centering
        \includegraphics[width=0.49\textwidth]{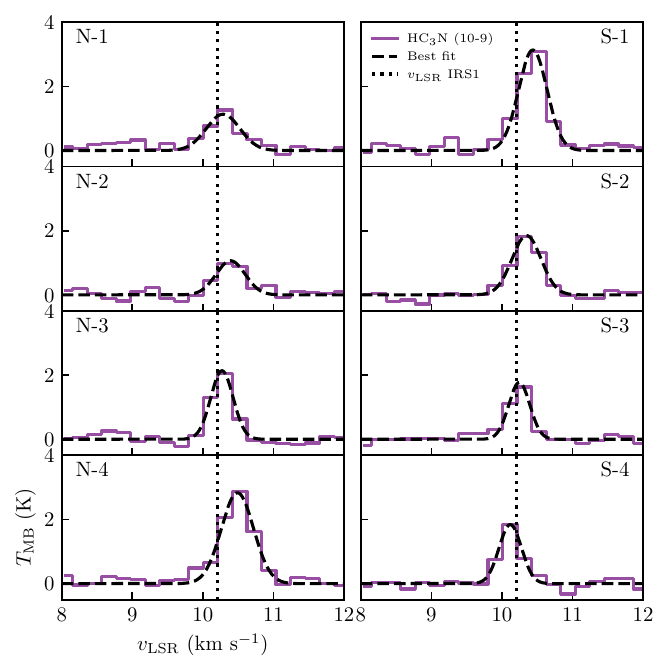}
        \caption{ \ce{HC3N}\,($10-9$) spectra at selected locations within both filaments, ordered in position from north to south. Fil-1 spectra, labeled N-X, are on the left column and Fil-2 spectra, labeled S-X, are on the right. Purple lines show the spectra, whereas black curves show the best fit Gaussian curve for each. }
        \label{fig:HC3Nspec}
\end{figure}

We fit one Gaussian profile to the \ce{HC3N}\,($10-9$) line emission using the Python package \texttt{pyspeckit} \citep{Ginsburg-Mirocha2011pyspeckit, Ginsburg2022pyspeckitpaper}. We describe our procedure in detail in Appendix \ref{ap:gaussfit}. In summary, we fit one Gaussian to all spectra with $S/N>5$ using moment \commentstere{maps} to estimate the initial parameters, and then attempt the fit again using the results as initial guesses. The best fit parameters of each spectrum are shown in Fig.~\ref{fig:HC3N-fitparams}. \commentstere{We select four locations within each filament, labeled N-1 to N-4 and S-1 to S-4 for Fil-1 and Fil-2, respectively, to show their spectra in Fig. \ref{fig:HC3Nspec}.} \refereecomment{The spectra are taken from a single pixel, which has a size of \refereetwo{(0.6\arcsec)$^2$}.} Almost all spectra with $S/N>5$ are well fit using only one Gaussian, as seen in Fig. \ref{fig:HC3Nspec}, with the exception of spectra located \commentstere{within a beam of} the protostar's position, which we explore further. 

At the position of B5-IRS1, there are \commentstere{several spectra within a radius of $ \sim 4\arcsec$} where it is possible to fit a second component to the \ce{HC3N}\,($10-9$) emission \commentstere{that has a lower velocity than its surroundings. We describe this second fit in more detail in Appendix \ref{ap:HC3Nsecondfit}}. Previous works have been able to detect two and even three Gaussian components in \ce{NH3} emission \citep[Choudhury et al. submitted,][]{Chen2022B5_NH3_vel}, so it is expected that \ce{HC3N}\,($10-9$) can trace more than one component as well. We describe how we evaluated this second fit in Appendix \ref{ap:gaussfit}. According to our criteria, there are not enough spectra with a good enough two Gaussian fit to justify keeping it in our analysis. To separate the gas emission into more components in this region, we require observations with higher spectral resolution, as the Gaussians are covered by 3 or 4 channels only in most cases \commentstere{(Fig. \ref{fig:HC3Nspec})}, which is not enough to fit two Gaussians with certainty. Therefore, we stay with the one Gaussian fit for all regions for further analysis.

The central velocities \vlsr of each spectrum, shown in Fig.~\ref{fig:HC3N-fitparams} center, range from 9.8 to 10.7 \kms. \refereecomment{The uncertainties in \vlsr are 0.025 \kms on average, ranging from 0.01 \kms at the center of the filaments and condensations, and increasing to approximately 0.05 \kms toward the edges of the filaments, where the $S/N\approx5$}. The \vlsr map shows \commentstere{filament-scale} gradients both \commentstere{parallel to} the direction of the filaments and perpendicular to them. The central velocity within Fil-1 shows a velocity gradient parallel to the filament length of approximately 0.4 \kms, starting from 10.3 \kms at the northern tip of Fil-1 and ending toward Cond-1 with 10.7 \kms. Perpendicular to \commentstere{this} filament, there is a gradient from lower to higher \vlsr from east to west, starting from 10.0 \kms outside of the filament defined in \ce{NH3} and reaching 10.4 \kms \commentstere{on the opposite side}. The gas within Fil-2 contour also shows a parallel velocity gradient, from 10.1 \kms in the south tip to 10.4 \kms in the north, reaching its maximum velocity within Cond-2 and 3. \commentstere{There is also a perpendicular gradient that is not apparent at first sight because it is smaller than the one present in} its northern counterpart. \commentstere{The largest observed difference between the edges of Fil-2 is at the southern edge of Cond-3, where the \vlsr goes from 10.4 \kms in the east to 10.2 \kms in the west.} The \ce{HC3N}\,($10-9$) velocity gradients are produced by two continuous bodies of extended emission, one spatially correlated to Fil-1 and another to Fil-2, as seen in the channel maps (Fig. \ref{fig:HC3N-chanmap}), and not from two or more spatially separated components in each filament. 

\commentstere{Within a beam} of the protostar there is a decrease in velocities \commentstere{with respect to its surroundings. At} Cond-2 and 3, the velocity is around 10.4 \kms, \commentstere{whereas} \commentstere{around B5-IRS1,} the velocity drops down to 10.0 \kms. \commentstere{This sudden blueshift is partially due to the second component \refereecomment{suggested} at that location. However, the main peak at the location of the protostar is still blueshifted with respect to B5-IRS1 \vlsr (Fig.~\ref{fig:ap-HC3N-protostarfit}). }

The \ce{HC3N}\,($10-9$) velocity dispersion stays rather constant along both filaments, varying within beam areas randomly between 0.12 and 0.2 \kms, except at the position of the protostar, where it increases suddenly to 0.5 \kms, and toward the south of Cond-3, where the dispersion increases to $\sim0.33$ \kms. This is the reason why the peak of integrated emission \refereecomment{(Fig. \ref{fig:HC3N-integrated})} is \refereetwo{at} the protostar's \refereetwo{location} but the peak in the $T_{\mathrm{peak}}$ map \refereecomment{(Fig. \ref{fig:HC3N-fitparams} left)} is located inside of Cond-2. The protostar might produce the higher velocity dispersion around it, including toward Cond-3. Assuming the gas temperature in this region is 9.7 K \citep[obtained from][]{Pineda2021B5ions}, the sound speed \refereecomment{is} $c_{\mathrm{s}} = 0.18$ \kms, so for the most part, the obtained velocity dispersion indicates the gas speed is mostly subsonic.

\subsection{\ce{H2CO} multicomponent line fitting \label{sec:h2co-fitting}}

\begin{figure*}[htbp]
        \centering
        \includegraphics[width=0.39\textwidth]{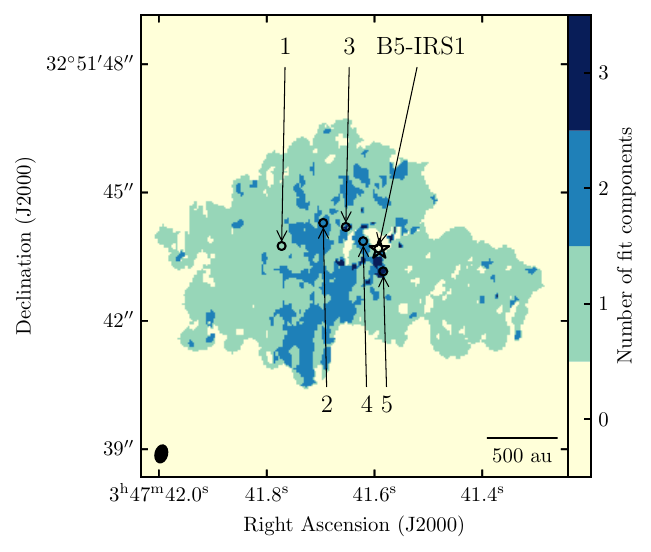}
        \includegraphics[width=0.59\textwidth]{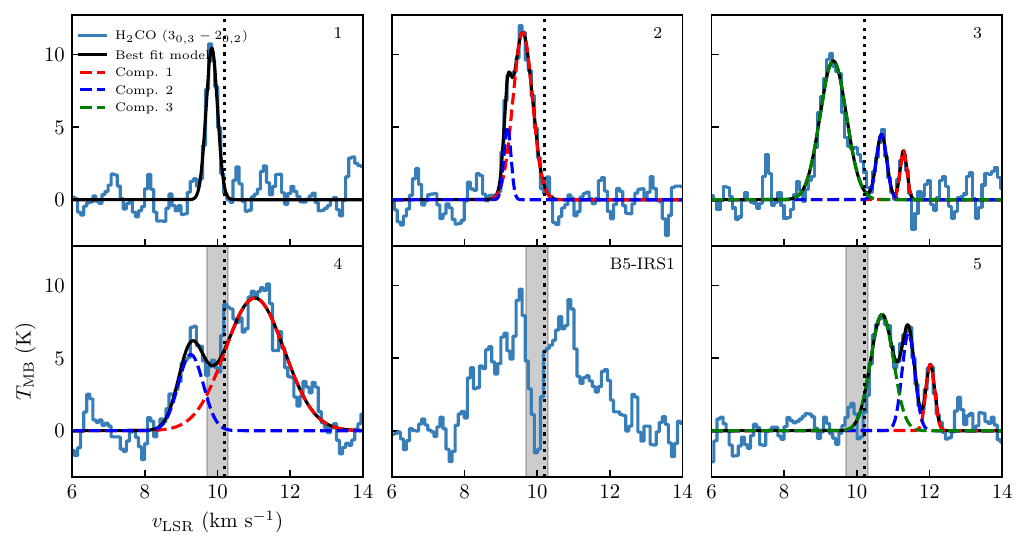}
        \caption{Results of the Gaussian fit to the ALMA \ce{H2CO} spectra. \commentstere{\textbf{Left:} }Number of components that best fit each \ce{H2CO} spectrum according to the criteria described in Appendix \ref{ap:gaussfit}. The black ellipse in the bottom left corner represents the beam size. The black star marks the position of the protostar.  \commentstere{Locations of sampled spectra are marked with empty black circles and labeled. \textbf{Right:} Selected \ce{H2CO} spectra with the best fit Gaussian components. \refereecomment{The spectra are taken from one single pixel.} The data are plotted with a solid blue line. The dashed lines represent the best fit Gaussians, with red for the first component, blue for the second and green for the third. The black solid line represents the sum of all components. The gray vertical area marks the channels that were masked for the fit for spectra located within 0\farcs8~ from the protostar. The dotted black vertical line marks the protostar's \vlsr. }
        }
        \label{fig:bestfitH2CO-ncomp}
\end{figure*}

\commentstere{Figure \ref{fig:bestfitH2CO-ncomp} shows a sample of spectra from the ALMA \ce{H2CO} emission cube, with increasing numbers representing \commentstere{decreasing} distance to the protostar. \refereecomment{The spectra are taken from a single pixel, which \refereetwo{has} an area of \refereetwo{(0.003\arcsec)$^2$}. }The spectra show that} close to and at the location of B5-IRS1,  \ce{H2CO} emission has more than one velocity component along the line-of-sight. \refereecomment{This component multiplicity explains the sudden increase of velocity dispersion shown in the moment 2 map (Fig. \ref{fig:H2CO-momentmap} right).}

We fit one, two, and three Gaussians to the \ce{H2CO} line emission for each spectrum inside a mask created using all pixels with $S/N>5$. The details of the procedure, including the mask creation and the criteria to determine the number of components along each line of sight, are detailed in Appendix \ref{ap:gaussfit}. The general procedure is similar to the Gaussian fit in Sect. \ref{sec:HC3Nfit}, but takes into account the possible effects of missing short- and zero-spacings in our interferometric data by masking the spectra between 9.7 and 10.3 \kms for pixels \commentstere{within a radius of 0\farcs8~ from the protostar. An observation of these central spectra determines that these are the most affected by missing short-spacing data.} 

The number of Gaussian components that best fit the spectra are \commentstere{shown} in Fig.~\ref{fig:bestfitH2CO-ncomp} left \commentstere{and we show spectra with their corresponding best fits in Fig.~\ref{fig:bestfitH2CO-ncomp}  right}. Most of the spectra are best fit using only one Gaussian, but closer to the protostar and toward the south, there are spectra that are better fit with 2 and 3 Gaussians. There are very few spectra that are best fit with 3 Gaussians, and \commentstere{most} of these are located in the region where we mask the central velocity channels (between 9.7 and 10.3 \kms) for the fit. Our criteria determined \commentstere{that in 0.1\% of the spectra, located at a distance  $<500$ au from the protostar, the best model has a probability lower than 95\% of representing a considerable improvement in comparison to the other models.} \refereecomment{In paritcular, at the position of the protostar, the spectrum shows two peaks with a dip at the \vlsr of the protostar. This dip is most probably caused by the missing short-spacing information, as it is located right on the channels that are masked. At this location, the Gaussian fits showed very large uncertainties for its parameters (more than 50\%), and therefore the spectra within one beam of the protostar are left unfit.}

\section{Analysis\label{sec:analysis}}

\subsection{Gas flow from Barnard 5 dense core to the filaments}

\subsubsection{Comparison between core \ce{HC3N} and \ce{NH3} velocities\label{sec:analysis-compHC3N-NH3}}

We compared the \ce{HC3N}\,($10-9$) central velocities \vlsr with the \ce{NH3}\,(1,1) \vlsr found in \citetalias{Pineda2015NatureB5cores} (shown in Fig.~\ref{fig:NH3-vlsr}) to observe the relative motions of the chemically fresh gas with respect to the filaments. \refereecomment{According to chemical models of cold, dark molecular clouds,  \ce{NH3} and other nitrogen-bearing molecules are regarded as a ``late-type" molecule, tracing advanced stages of gravitational core collapse, whereas \ce{HC3N} and the cyanopolyyne (HC$_{2n+1}$N) molecule family trace chemically fresh gas \citep[i.e., unprocessed by core collapse processes,][]{Herbst-Leung1989carbon-chain-chem, Bergin-Tafalla2007, Aikawa2001chem-prestellar-cores}. }
We subtracted the \ce{NH3} \vlsr \commentstere{from the} \ce{HC3N}\,($10-9$) \vlsr, obtaining $\delta$\vlsr $= v_{\mathrm{LSR, HC}_3\mathrm{N}} - v_{\mathrm{LSR, NH}_3}$. 
For this, we first convolved the \ce{HC3N}\,($10-9$) \vlsr (Fig. \ref{fig:HC3N-fitparams} Center) to the spatial resolution of the \ce{NH3}\,(1,1) data (Table \ref{tab:obsprops}), using the \texttt{photutils} routine \texttt{create\_matching\_kernel} to obtain the matching kernel between the image beams and the \texttt{astropy.convolution} function \texttt{convolve} to smooth the map. 
Afterwards, we reprojected the spatial grid of the \ce{NH3}\,(1,1) central velocities image to the \ce{HC3N}\,($10-9$) spatial grid, using the \texttt{reproject\_exact} routine from the \texttt{reproject} python package. We regridded the \ce{NH3} map to the \ce{HC3N}\,($10-9$) map grid because the pixel size of the \ce{NH3} map is about 1.5 times smaller than the \ce{HC3N}\,($10-9$) image pixel size. Finally, we subtracted the \vlsr of \ce{NH3} \commentstere{from the  \ce{HC3N}\,($10-9$)} \vlsr. 

\begin{figure*}[htbp]
        \centering
        \includegraphics[width=0.51\textwidth]{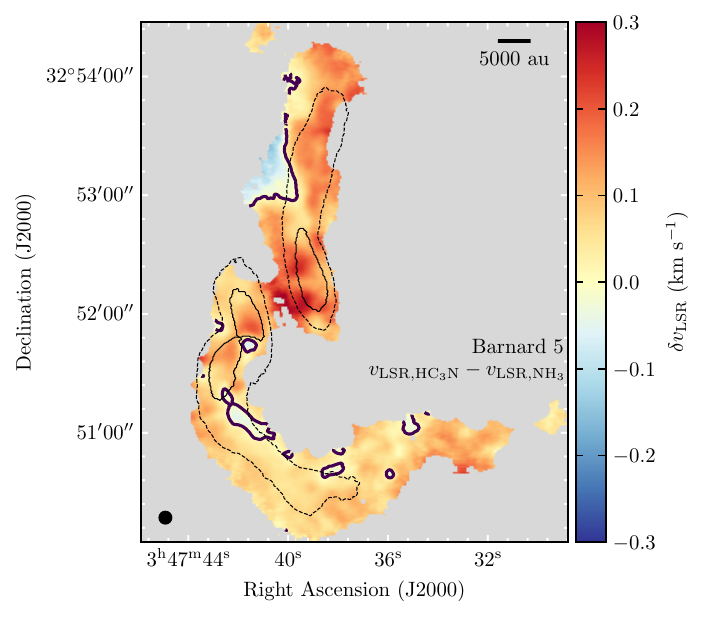}
        \includegraphics[width=0.47\textwidth]{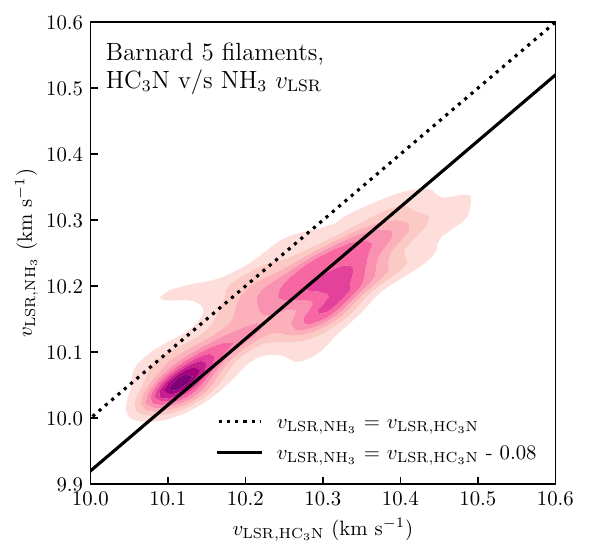}
        \caption{Difference between \ce{HC3N}\,($10-9$) line emission and \ce{NH3}\,(1,1) line emission central velocities in both filaments of Barnard 5. Left: Spatial map showing the difference of \ce{HC3N}\,($10-9$) central velocities with respect to \ce{NH3}\,(1,1) velocities. Black \commentstere{thick} contours mark the positions where the difference is 0. The resolution of the \ce{HC3N}\,($10-9$) data is on the lower left corner. The dotted contours show the filaments \commentstere{and the black solid contours outline the} \refereetwo{condensations} as defined in \citetalias{Pineda2015NatureB5cores} and \citetalias{Schmiedeke2021supercriticalfil}. Right: v$_{\mathrm{LSR}, \mathrm{HC}_3\mathrm{N}}$ versus v$_{\mathrm{LSR}, \mathrm{NH}_3}$ Distribution of the velocity difference shown using \refereecomment{its KDE}. The dashed vertical line represents the 1:1 line, and the solid line is shifted by -0.08, which is the median value of the difference between both velocities. }
        \label{fig:HC3N-NH3}
\end{figure*}

Figure \ref{fig:HC3N-NH3} left shows the resulting $\delta$\vlsr in the image plane. The map shows that \ce{HC3N}\,($10-9$) is \refereecomment{consistently} redshifted with respect to \ce{NH3} except for a few locations, mostly without systematic variations within the \ce{NH3} filament regions. The median value of $\delta$\vlsr is $0.08^{+0.06}_{-0.05}$~\kms.  \refereecomment{This difference is more than 3 times larger than the average uncertainty for $\delta$\vlsr, which is dominated by the uncertainty of the \ce{HC3N}\,($10-9$) \vlsr (Sect. \ref{sec:HC3Nfit}).} Toward the southern tip of Cond-1, \ce{HC3N}\,($10-9$) \vlsr is larger than \ce{NH3}\,(1,1) \vlsr by $\approx0.28$~\kms, but the rest of the spectra are consistent with the median value. Figure \ref{fig:HC3N-NH3} Right shows the \ce{HC3N}\,($10-9$) central velocities versus the respective \ce{NH3} \vlsr at the same location, estimated using a two-dimensional Gaussian Kernel Density Estimate (KDE) obtained with the \texttt{scipy.stats} python package \texttt{gaussian\_kde} function \citep{2020SciPy-NMeth} assuming all pixels have the same weight. The KDE shows there is an approximately linear relation between both \vlsr with a shift of $-0.08$~\kms. This comparison shows that \ce{HC3N}\,($10-9$) follows a different kinematic component of the gas in Barnard 5 than \ce{NH3}.

There are a few locations where \ce{HC3N}\,($10-9$) emission is blueshifted with respect to \ce{NH3}\,(1,1) emission. At the right of Fil-1, $\delta$\vlsr drops down to $\approx-0.1$ \kms. Also, just below Cond-3 in Fil-2, there is a region where $\delta$\vlsr also reaches $\approx-0.1$. The blueshifted emission at these positions, which are elongated along the filaments, suggest there is a perpendicular velocity gradient of \ce{HC3N} with respect to \ce{NH3}. \commentstere{In particular}, there is a sudden shift from positive to negative $\delta$\vlsr at the location of the protostar, reaching down to $\delta$\vlsr $\approx -0.06 \kms$. The area of the reversal is approximately the \ce{NH3}\,(1,1) beam size. \commentstere{Comparing the \ce{HC3N}\,($10-9$) fit with the \ce{NH3}\,(1,1) fit (Fig.~\ref{fig:NH3-vlsr})} shows that both molecules present a sudden decrease in \vlsr at that location, but the difference between \vlsr in this position and its surroundings is larger for  \ce{HC3N}\,($10-9$) than for \ce{NH3}.

\subsubsection{\ce{HC3N} velocity gradients \label{sec:grad}}

In Fig. \ref{fig:HC3N-fitparams} center, there is a velocity gradient perpendicular to Fil-1 observed in \ce{HC3N}\,($10-9$) central velocities. \commentstere{The} perpendicular gradient in Fil-2 \commentstere{is overshadowed by} a strong gradient running parallel to the filament length. We calculated the velocity gradients  $\nabla$\vlsr present in the map to determine \commentstere{the direction of gas flow at the positions where the condensations and the protostar are and to determine} if there are gradients perpendicular to Fil-2 that are hidden by an analysis by eye. We \refereecomment{employed} the same procedure used to calculate velocity gradients in Barnard 5 \ce{NH3}\,(1,1) emission by \cite{Chen2022B5_NH3_vel}, which is described in detail in \cite{Chen2020NGC1333fils}. In summary, we calculated the gradient pixel-by-pixel by fitting a plane centered around the pixel, using a square aperture with a width of \commentstere{2} beams ($\approx 8\arcsec$ width), to ensure the capture of velocity gradients across non-correlated regions. We only fit pixels which have a number of neighboring pixels equivalent to one third of the aperture area available to ensure the quality of the fit. 

\begin{figure}[htbp]
    \centering
    \includegraphics[width=0.49\textwidth]{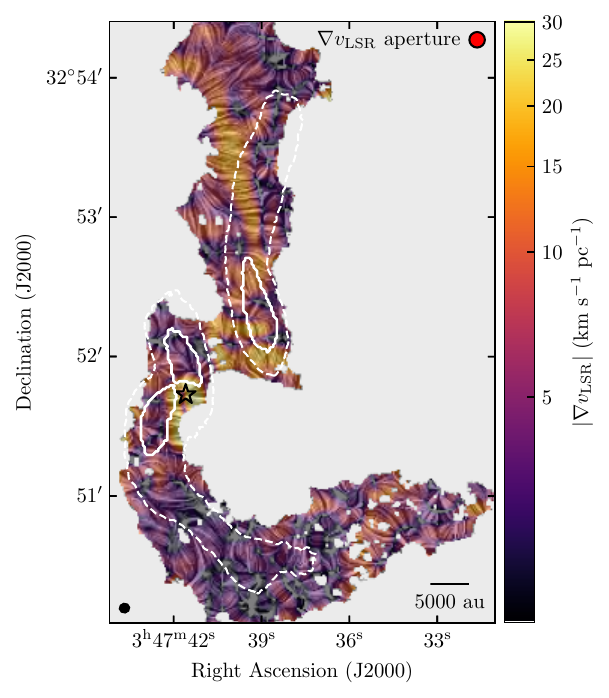}
    \caption{Velocity gradient \commentstere{intensities} $|\nabla\vlsr|$ calculated from the Gaussian fit to \ce{HC3N}\,($10-9$) emission, \commentstere{with the gradient directions} visualized \commentstere{as an overplotted texture} using line integral convolution \citep[LIC,][]{Cabral+Leedom1993LIC}. The black star indicates the position of B5-IRS1. The black ellipse in the bottom left corner represents the beam size.  \commentstere{The red circle at the top right corner represents the area used to calculate the gradients. White dashed contours outline the \ce{NH3} filaments \citepalias{Pineda2015NatureB5cores, Schmiedeke2021supercriticalfil}. White solid contours outline the edges of the condensations \citepalias{Pineda2015NatureB5cores}. The scalebar at the bottom right corner represents a physical distance of 5000 au.}}
    \label{fig:HC3N-nablav}
\end{figure}

Figure \ref{fig:HC3N-nablav} shows the resulting \ce{HC3N}\,($10-9$) $\nabla$\vlsr visualized using line integral convolution \citep[LIC,][]{Cabral+Leedom1993LIC} textures over the absolute magnitude of the gradient $|\nabla\vlsr|$. We used the \texttt{LicPy} package\footnote{https://github.com/drufat/licpy} \citep{Rufat2017} to \commentstere{generate a texture that represents the directions of the gradients,} and then overlaid the resulting texture. \commentstere{The uncertainty in $|\nabla\vlsr|$ is between 0.3 and 1  \kms pc$^{-1}$ and increases toward the edges of the \ce{HC3N}\,($10-9$) emitting region.} We focus our discussion on a qualitative description of the resulting $\nabla$\vlsr map. 

The vector field is well ordered within the brightest \ce{HC3N} emissions, in particular, in and around the condensations and filaments defined by \ce{NH3} contours and toward the bright emission toward the north of the map \citepalias[which connects with Clump-1 from][]{Schmiedeke2021supercriticalfil}. Figure \ref{fig:HC3N-nablav} reveals that gas in and around both filaments have both parallel and perpendicular gradients. 

\commentstere{The $\nabla$\vlsr field reveals the flow of gas toward the condensations.} Within the contours of Fil-1, the perpendicular gradients are toward the edges of \ce{HC3N} emission with $|\nabla\vlsr|\,\approx 12 - 15$ \kms pc$^{-1}$). The vector field reveals a strong (\refereetwo{|}$\nabla$\vlsr $|\,\approx 20$  \kms pc$^{-1}$ ) parallel gradient of $\sim 8000$ au in length above Cond-1.  The vector field is particularly well ordered at the north of Fil-1, where there is \ce{HC3N} emission which could possibly connect to Clump-1 toward the north. \commentstere{Cond-1 is surrounded by flows with magnitudes between 25 and 30 \kms pc$^{-1}$. that point toward its center, but at the center of the condensation the flow stagnates (i.e., drops to 0 \kms pc$^{-1}$). This indicates \ce{HC3N} moves inward to Cond-1 from all directions.}

\begin{figure}[htbp]
        \centering
        \includegraphics[width=0.49\textwidth]{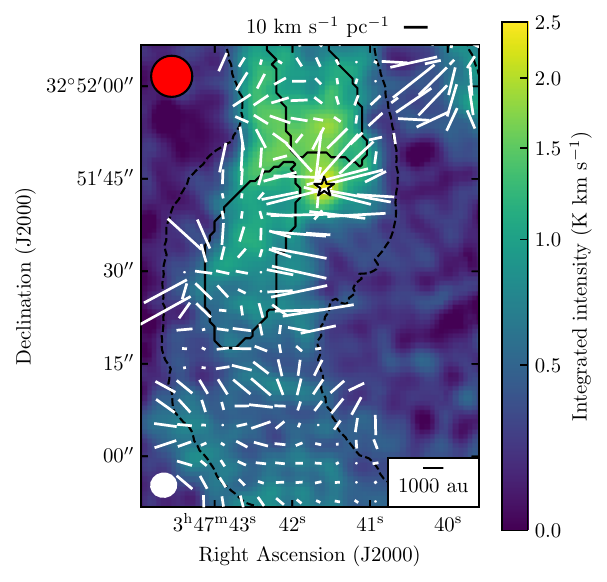}
        \caption{\commentstere{Zoom of the integrated intensity map toward Cond-2, -3 and the protostar, with the  $\nabla$\vlsr vector \refereecomment{orientations}} from Fig.~\ref{fig:HC3N-nablav} shown with white \refereecomment{lines}. The black \refereecomment{line} at the top represents a gradient magnitude of 10 \kms pc$^{-1}$. The black star indicates the position of B5-IRS1. The white ellipse in the bottom left corner represents the beam size. Black dashed contours outline the \ce{NH3} filaments \citepalias{Pineda2015NatureB5cores, Schmiedeke2021supercriticalfil}. Black solid contours outline the edges of the condensations \citepalias{Pineda2015NatureB5cores}. \refereetwo{The red circle at the top left corner represents the area used to calculate the gradients.} The scale bar at the \refereecomment{bottom} right marks a length of 1000 au.}
        \label{fig:HC3N-gradient-zoom}
\end{figure}

The $\nabla$\vlsr calculation also reveals the parallel flow of gas in Fil-2 that was not evident by eye. Toward the south of Cond-3, most of Fil-2 shows perpendicular gradients with $|\nabla v_{\mathrm{LSR}}|$ between 3 and 10 \kms pc$^{-1}$, \commentstere{much lower than the perpendicular gradients at Fil-1}. The western edge of Cond-3 presents the strongest velocity gradient of the map ($\nabla v_{\mathrm{LSR}} \approx 40$  \kms pc$^{-1}$) and its direction is parallel to the filament. \commentstere{Both condensations (Cond-2 and -3) show stagnating flows within them. This indicates that gas flows from the outside in, as in the case of Cond-1.} Most notably, \commentstere{a zoom into the condensations within Fil-2, shown in Fig.~\ref{fig:HC3N-gradient-zoom}, reveals that} the gradient  \commentstere{direction} \commentstere{curves toward the protostar} and the parallel gradient ends there as well.  This plot suggests that fresh gas, traced by \ce{HC3N}, flows toward the protostar, \commentstere{and the large-scale flow along the filament is hence shared by the protostar, Cond-2 and Cond-3.}

 \subsubsection{Mass inflow from Barnard 5 dense core to the filaments}
 
\commentstere{We estimated the mass transport and the infall rate from the flows traced by  \ce{HC3N} emission. In Choudhury et al. (submitted), they estimate an infall rate $\dot{m}$ for Fil-1 of $1.1\times10^{-4}$ \Msun yr$^{-1}$ and for Fil-2 of $1.8\times10^{-4}$ \Msun yr$^{-1}$. We used the same geometry for both filaments and their Eq. 8 to estimate $\dot{m}$:}
\begin{equation}
        \dot{m} = S \rho v_{\mathrm{inf}},
 \end{equation}
\commentstere{where $S$ is the surface area where infall occurs, $\rho$ is the total mass volume density within the surface area and $v_{\mathrm{inf}}$ is the estimated infall velocity. }

We obtained $\rho = n(\mathrm{H}_2)\mu m_{\mathrm{H}}$ by using the line ratio between \ce{HC3N}\,($10-9$) and ($8-7$) molecular transitions. This procedure is described in detail in Appendix \ref{sec:coremass}. In summary, we obtained a number density $n(\mathrm{H}_2)\approx10^5$ cm$^{-3}$ based on the non-LTE radiative transfer code RADEX \citep{vanderTak2007RADEX}. The mean difference in velocity, which is defined as half the difference in \vlsr across a filament, is $v_{\mathrm{inf}}=0.2$ \kms for Fil-1 (the same as the one observed in Choudhury et al. submitted) and 0.1 \kms for Fil-2 (Sect. \ref{sec:HC3Nfit}). The resulting mass infall rate is $\dot{m} = 4.3\times10^{-5}$ \Msun yr$^{-1}$ in Fil-1 and $\dot{m} = 3.5\times10^{-5}$ \Msun yr$^{-1}$ \commentstere{for Fil-2}, about 3 times lower than the Fil-1 infall rate found in Choudhury et al. (submitted). \refereetwo{The difference is due to the obtained $n(\mathrm{H}_2)$ in both works. We estimated $n(\mathrm{H}_2)$ using our observed \ce{HC3N} transitions (Appendix \ref{sec:coremass}). On the other hand, Choudhury et al. (submitted) used the filament masses from \citetalias{Schmiedeke2021supercriticalfil}, which in turn were calculated by comparing Barnard 5's \ce{NH3} emission and the JCMT 0.45 mm emission.}

\subsection{Infall from envelope to disk scales}

\subsubsection{Clustering of physically related structures in \ce{H2CO}\label{sec:dbscan-clustering}}

We separated the different Gaussian components found in \ce{H2CO} in Sect. \ref{sec:h2co-fitting} into clusters, which helps to interpret the kinematic properties of the B5-IRS1 envelope. It is \commentstere{challenging} to do this separation by eye, especially where there are two components in the same line of sight that can belong to two different components indistinguishably. For example, in the boundary between blueshifted and redshifted sides of the \ce{H2CO} emission, toward the south \refereecomment{and east} of the protostar, there are two Gaussian components with velocities blueshifted with respect to the protostar's \vlsr(see Fig.~\ref{fig:bestfitH2CO-ncomp} \refereecomment{Left}). \commentstere{Also, the locations where there are three Gaussian components are small and sparse, as well as spectra at the east and west edges of the \ce{H2CO} emission. It is possible that the additional fitted components are either noise or extended emission that is not fully sampled due to missing short- and zero-spacing observations.} \refereecomment{We describe our procedure in Appendix \ref{ap:clustering}, where we also show the resulting clusters in Fig. \ref{fig:dbscan-total}.} In summary, we employed the density-based spatial clustering of applications with noise (DBSCAN) algorithm to separate the different physical components in position-position-velocity space.

\begin{figure}[htbp]
    \centering
    \includegraphics[width=0.49\textwidth]{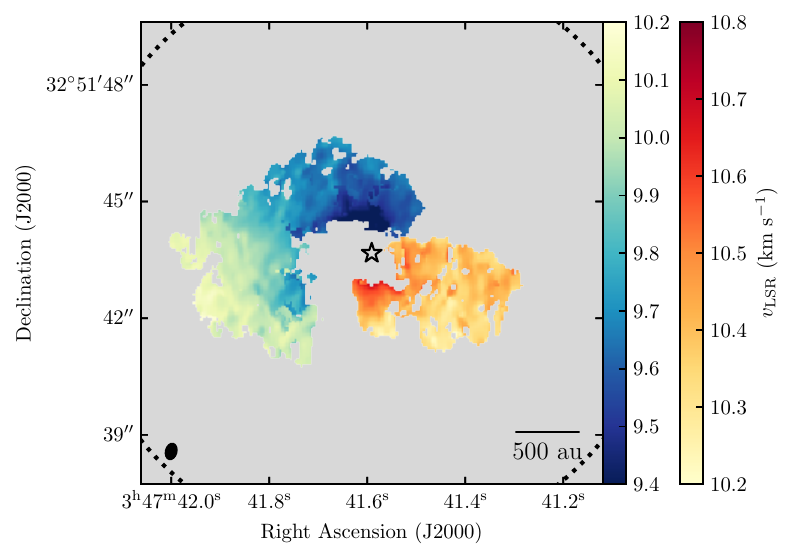}
    \caption{Central velocities for the blueshifted and redshifted \refereecomment{clusters} found using \refereecomment{the methods described in Appendix \ref{ap:clustering}}. The colorbar from yellow to blue corresponds to the velocities for blueshifted \refereecomment{cluster}, whereas from yellow to red corresponds to the redshifted \refereecomment{cluster}. The black ellipse in the bottom left corner represents the beam size. The black star represents the position of the protostar. The black dashed circle \refereetwo{(only seen at the corners)} represents the size of the primary beam. We note that the velocity ranges for both envelope components are not symmetric.}
    \label{fig:dbscan-redblue-velocities}
\end{figure}

The two largest clusters returned by DBSCAN (\refereecomment{Groups 0 and 1 from Fig. \ref{fig:dbscan-total}}) contain about 90\% of the points not categorized as noise. \refereecomment{The largest cluster is a group of points that are all blueshifted with respect to the protostar's \vlsr(10.2 \kms), and the second largest, a fully redshifted group.} We refer to these two groups as blueshifted and redshifted \refereecomment{clusters} throughout the rest of this work. \refereecomment{More details about these two clusters are in Appendix \ref{ap:clustering}}. The central velocities of the Gaussian fits belonging to each of these \refereecomment{clusters} are shown in Fig. \ref{fig:dbscan-redblue-velocities}.
The blueshifted \refereecomment{cluster} has a velocity gradient toward lower (more blueshifted) velocities as it gets closer to the protostar. This gradient is stronger for higher declinations. The redshifted  \refereecomment{cluster} has an increase in velocity when it gets closer to the protostar. 

The remaining three clusters represent $\approx10$\% of the points not considered noise. \refereecomment{The clustering algorithm is able to assign one of the two blueshifted Gaussian components toward the south of the protostar (Fig. \ref{fig:bestfitH2CO-ncomp}) to the blueshifted cluster and leave the other as another group (Group 3 from Fig. \ref{fig:dbscan-total}). This allows us to observe the velocity gradient more clearly. The main difference between the largest clusters and groups 4 and 5 is these two last groups show highly blueshifted and redshifted \vlsr very close to the protostar, representing a jump in the velocity gradient of groups 0 and 1. Group 3 has a central velocity of 10.8 \kms and group 4, of 9.2 \kms. Groups 3, 4, and 5,} as well as possibly part of the points categorized as noise, could be emission coming from other mechanisms observed in \ce{H2CO}, such as inner gas disk rotation or a more extended envelope, that due to the resolution or the missing short-spacing data cannot be fully observed. \refereecomment{Group 3, for example, shows a central velocity redshifted with respect to the protostar at the same position as the redshifted high-velocity \ce{C^{18}O} emission, described in Appendix \ref{ap:gaussfit}. We suggest the high-velocity \ce{C^{18}O} and  \ce{H2CO} emission \refereetwo{components} trace the protostellar disk kinematics.} Another possibility is that within 500 au from the protostar \ce{H2CO} gas is optically thick, which would cause self-absorbed emission  that is being fitted as separate velocity components and then categorized as \commentstere{the purple and orange groups (labeled as 3 and 4, respectively)} by the algorithm. 

\subsubsection{Streamline model of the \ce{H2CO} clusters\label{sec:slmodel}}

The two main clusters found by DBSCAN each have a gradient where velocity (with respect to the protostar) increases as distance decreases, similar to streamers confirmed kinematically toward other protostars, for example Per-emb-2 \citep{Pineda2020Per2streamer} and Per-emb-50 \citep{Valdivia-Mena2022per50stream}. We determined if the kinematics of the blueshifted and redshifted components are consistent with streamers infalling toward the protostellar disk of B5-IRS1.

First, we checked if \ce{H2CO} emission is tracing or is affected by the outflow. In some protostellar sources, \ce{H2CO} traces the outflow or the extremely high-velocity jet \citep{Tychoniec2021moltracers}. The position of the blueshifted component is at the east of the protostar, in the same side as the blueshifted outflow cone, and the redshifted component and outflow cone are both toward the west (Fig.~\ref{fig:H2CO-integrated}). However, the velocity gradients of the blueshifted and redshifted components are opposite to the gradient expected for an outflow. For the blueshifted component, \vlsr goes from 10.1 \kms at a distance of about 1000 au to 9.2 \kms at approximately 200 au from the protostar. This means that it is accelerating toward more blueshifted velocities with respect to the protostar's \vlsr (10.2 \kms) as distance decreases. For the redshifted component, \refereecomment{\vlsr is approximately} 10.3 \kms at about 1000 au and accelerates to 10.6 \kms at a distance of 300 au, which also \refereecomment{represents} an acceleration with respect to the protostar' \vlsr with decreasing distance. For an outflow, the velocity with respect to the protostar increases with distance, opposite to the behavior of \vlsr shown by both components. Therefore, the motion observed in these \ce{H2CO} components is not consistent with outflow motion.

Second, we checked the velocity dispersion $\sigma_{\mathrm{v, fit}}$ of the Gaussians in the blueshifted and redshifted \refereecomment{clusters} and compare them to the thermal sound speed. Nonthermal velocity dispersion larger than the sound speed can indicate that the motion traced by \ce{H2CO} is not simple envelope infall, but can be affected by turbulence and shocks produced by the outflow in the same line of sight. 
We describe the procedure and the resulting nonthermal velocity dispersion of the blueshifted and redshifted clusters in Appendix \ref{ap:nonthermalH2CO}. The velocity dispersion is subsonic for the majority of the emission in both components, but it becomes trans-sonic for emission within $r<600 $ au from the protostar. This suggests that within this radius,  \ce{H2CO} gas is possibly affected by the outflow.
\refereecomment{Additionally, there is a larger density of trans-sonic emission for the redshifted component. This is due to the smaller area covered by this component and the fact that it is mostly contained in the observed direction of the outflow cone. The \ce{H2CO} gas contained in this cluster is therefore contaminated by outflow emission. For this reason, we leave the redshifted component as a streamer candidate and do not do further analysis on it in this work.}

We modeled the kinematics of the \refereecomment{blueshifted cluster} observed with H$_2$CO emission to confirm that the velocity gradient observed in Fig. \ref{fig:dbscan-redblue-velocities} is consistent with streamer motion, using the analytic solution for material falling from a rotating, finite-sized cloud toward a central object, dominated by the gravitational force of the latter. We used the analytic solutions of \cite{Mendoza2009}, previously used by \cite{Pineda2020Per2streamer} and \cite{Yen2014L1489Infall}. The model's input and output are described in detail in \cite{Valdivia-Mena2022per50stream}, here we describe briefly the input parameters used for this source.

The streamline model requires the central mass that is causing the gravitational pull as input, which is the sum of the masses of the protostar, disk, and envelope, $M_{\mathrm{tot}}=M_{*}+M_{\mathrm{disk}}+M_{\mathrm{env}}$. We estimate the protostellar mass by fitting a two-dimensional Gaussian to individual \ce{C^{18}O} (2 -- 1) channels and comparing the resulting distance and velocity (with respect to B5-IRS1) with the Keplerian curves predicted for different protostellar masses. We describe the process in detail in Appendix \ref{sec:starupperlim}. Through this analysis, we obtain a central protostellar mass $M_{*}\approx0.2$ \Msun. \commentstere{For the disk mass, we use \refereecomment{an upper limit of} $M_{\mathrm{disk}}=0.03$ \Msun found by \cite{Zapata2014B5outflow}. \refereecomment{As this value is small, it does not have a significant effect on the resulting parameters of the model}}. \refereetwo{We use an envelope mass of $M_{\mathrm{env}}=0.27$ \Msun from \cite{Andersen2019mass}}, \refereecomment{obtained through the comparison of continuum observations from the Submillimeter Array (SMA) and single-dish observations. The disk and envelope mass are corrected for a distance to B5 of 302 pc \citep{Zucker2018distance}, as the original masses reported in those works are calculated with a distance of 240 pc.}

To decrease the number of free parameters to explore, we fixed the inclination angle $i$ and position angle $PA$ of the rotating cloud using previous information from the outflow and disk in B5-IRS1. The outflow inclination angle is -13\degr \citep[where 0\degr is on the plane of the sky and positive is away from the observer,][]{Yu1999B5outflow} and its position angle is 67\fdg1 \citep[where 0\degr is north,][]{Zapata2014B5outflow}. Assuming the disk belongs in a plane perpendicular to the outflow, then $PA=157\fdg1$ (from north, if 0\degr is a disk aligned in the north to south direction) and $i=13\degr$ (from the plane of the sky). In this setup, the angular velocity vector of the disk $\vec{\omega}$, and therefore of the streamer, points toward the southwest, away from the observer.

\begin{figure*}[htbp]
    \centering
    \includegraphics[width=0.55\textwidth]{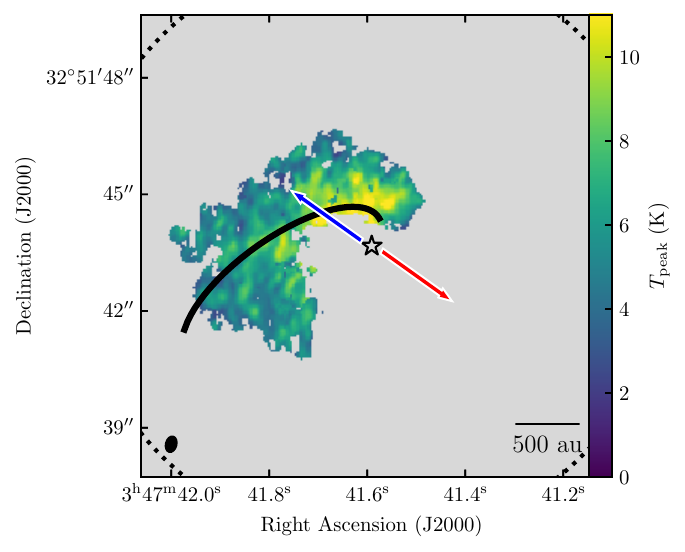}
    \includegraphics[width=0.44\textwidth]{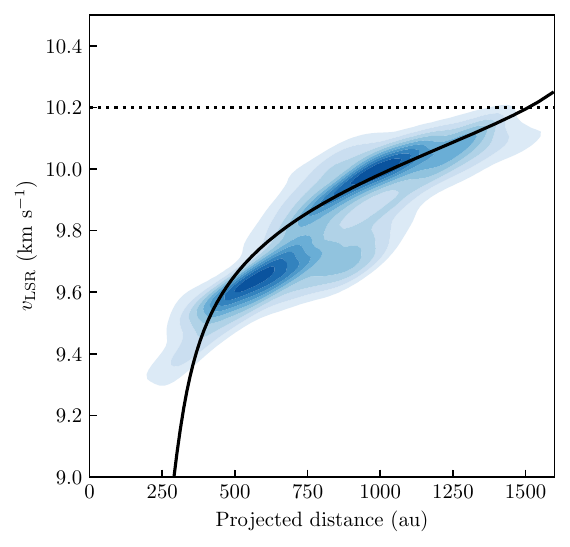}
    \caption{Streamline model that best fits the observed blueshifted envelope component in H$_2$CO line emission. In both image and velocity space, the black curve represents the streamline model solution that best describes the bulk motion of the streamer. 
    \textbf{Left:} Peak main beam temperature $T_{\mathrm{peak}}$ for each pixel in the blue component found with DBSCAN. The blue and red arrows show the direction of the blue and red outflow \refereetwo{lobes} from \cite{Zapata2014B5outflow}, respectively.  \refereetwo{The black ellipse in the bottom left corner represents the beam size. The black star represents the position of the protostar. The black dashed circle represents the size of the primary beam. } \textbf{Right:} central velocity KDE for the blueshifted envelope plotted against the projected distance from the protostar.}
    \label{fig:sl-model}
\end{figure*}

We explored the initial position ($r_0$, $\vartheta_0$ and $\varphi_0$) and velocity $\mathrm{v}_{r,0}$ to find the set that best fits the observed positions and velocities. We manually changed the parameters to obtain streamline model curves that resemble the shapes of both the peak emission and the line-of-sight velocities of each component (Fig.~\ref{fig:sl-model}). 
As the infall motion covers a large area in the image plane and does not look like a thin, collimated streamer \citep[as in the cases of][]{Pineda2020Per2streamer,Valdivia-Mena2022per50stream}, there must be a family of solutions that describe the whole streamer. We find the streamline model that best describes the bulk motion of the blue cluster, that is, the one that best fits the observed velocities and is contained within the emission region of the blue cluster in the image plane. Its parameters are in Table \ref{tab:paramsstream}. \commentstere{We explored other combinations of $r_0$, $\vartheta_0$ and $\varphi_0$ to find other possible solutions. The parameters presented in Table \ref{tab:paramsstream} are the ones that replicate \refereecomment{best} the approximate shape of the streamer in the image plane and the velocities at the same time. } 

Figure \ref{fig:sl-model} shows the projected trajectory of the streamline model over the blue \refereecomment{cluster} peak temperature (left panel) and over the KDE of the velocities and projected distance in the streamer (right panel). We used the KDE implementation in the python package \texttt{SciPy} \citep{2020SciPy-NMeth} over the resulting central velocities obtained for \refereecomment{the} cluster. The streamline model reproduces the velocity gradient shown in the KDE. The resulting centrifugal radius $r_\mathrm{c}$ \citep[as defined in][]{Mendoza2009}, which is the distance down to which the movement can be modeled with constant angular momentum, is 249 au. \refereecomment{This radius is an upper limit to the location where the mass flow ends, as any process that can affect the streamer motion, such as interaction with the outflow (as mentioned in Appendix \ref{ap:nonthermalH2CO}), will decelerate the gas flow and therefore, reduce its angular momentum.} This radius is consistent with the infall material being deposited at \refereecomment{gas} disk-scale distances from the protostar: \refereecomment{the dust disk has an estimated radius $\sim50$ au using mm-continuum emission \citep{Yang2021PEACHES}, and although we cannot define a gas disk radius with our \ce{C^{18}O} data (Appendix \ref{sec:starupperlim}), the gas disk extension tends to be much larger than what can be seen in dust emission, up to several 100 au \citep[][and references within]{Miotello2022PPVII}.}

\begin{table}
        \caption{\label{tab:paramsstream}Parameters of the streamline model that reproduce best the H$_2$CO blueshifted \refereecomment{cluster} and resulting centrifugal radius.}
        \centering
        \begin{tabular}{ccc}
                \hline\hline
                Parameter  & Unit   & Value     \\ \hline
                $\vartheta_0$ & deg   & 101  \\
                $\varphi_0$   & deg    & 65   \\
                $r_0$      & au     &   2810 \\
                $v_{r,0}$  & \kms   & 0  \\
                $\Omega_0$ & s$^{-1}$ & \refereetwo{$2.8 \times 10^{-13}$}  \\
                $i$        & deg   & 13    \\
                P.A.       & deg  & 157.1   \\ \hline
                $r_\mathrm{c}$ & au & \refereetwo{245} \\ \hline
        \end{tabular}
\end{table}

\section{Discussion\label{sec:discussion}}

\subsection{Chemically fresh gas feeds the filaments\label{sec:disc-freshgas}}

In this work, we present new \ce{HC3N}\,($10-9$) and ($8-7$) emission data within the coherent core of B5. Our results and analysis of \ce{HC3N}\,($10-9$) velocities and its comparison to previous results with \ce{NH3}\,(1,1) emission indicate that \ce{HC3N} traces chemically fresh gas infalling toward the filaments \commentstere{and condensations}, following the \commentstere{curvature of the} filaments. We explain our reasoning below.

Figure \ref{fig:HC3N-NH3} shows that \ce{HC3N} emission is consistently redshifted with respect to \ce{NH3}. \commentstere{The systematic difference in centroid velocities shows} these molecules do not trace the same material within B5. \refereecomment{As mentioned in Sect. \ref{sec:analysis-compHC3N-NH3}, \ce{HC3N} traces material unprocessed by core collapse, whereas \ce{NH3} is expected to be more abundant in regions where core collapse is underway.} \commentstere{Previous results in other star formation regions also show kinematic differences between carbon-chain molecules and nitrogen bearing molecules \citep[e.g.,][]{Friesen2013infall-serpens-south}. }  \commentstere{Therefore, \ce{HC3N} traces the kinematics of chemically fresh material \refereecomment{with respect to the main filamentary structure}.} 

That being said, the velocities shown by \ce{HC3N} emission are consistent with infall from the \commentstere{B5 dense} core to the filaments \commentstere{and condensations}. \commentstere{The Gaussian fit to the \ce{HC3N} spectra (Fig.~\ref{fig:HC3N-fitparams}) \refereecomment{and the further velocity gradient analysis (Sect. \ref{sec:grad})} reveal that both filaments show a velocity gradient running parallel to the filaments' full length. \commentstere{The velocity gradients of the filaments run in opposite directions}: in Fil-1, the global gradient runs from lower \vlsr toward the north to higher \vlsr in the south, with the highest \vlsr in Cond-1, whereas in Fil-2 the global gradient starts from lower \vlsr at the south and increases toward the north, following the curvature of the filament.} We interpret this as a flow of gas running along the direction of the filaments toward the condensations. \commentstere{\ce{NH3}\,(1,1) emission also shows parallel velocity gradients with the same orientations \citepalias[Fig.~\ref{fig:NH3-vlsr},][]{Schmiedeke2021supercriticalfil}. These large-scale parallel velocity gradients are similar in to those found in other filament, such as NGC 1333 \citep[][]{Hacar2017NGC1333Fibers,Chen2020NGC1333fils} and L1517 \citep{Hacar-Tafalla2011L1517}.} \commentstere{Assuming filament lengths of 0.24 pc for Fil-1 and 0.31 for Fil-2 from \citetalias{Schmiedeke2021supercriticalfil}, the average gradients along each filament are approximately 1.7 \kms pc$^{-1}$ and 1.0  \kms pc$^{-1}$, respectively. These values resemble the magnitude of the parallel velocity gradients found in other filaments and fibers \citep[$\sim1$ \kms pc$^{-1}$, e.g.,][]{ Kirk2013Serpens-south,FernandezLopez2014CLASSy-serpensfilaments,Chen2020NGC1333fils}. }

\commentstere{The velocity gradients running perpendicular to the filament orientations are suggestive of infall from the B5 dense core to the filaments themselves.} Fil-1 and the \ce{HC3N} emission just outside of it to the north and to the east \commentstere{show an east-west velocity gradient that crosses the full width of the filament (Fig.~\ref{fig:HC3N-fitparams} center)} whereas \commentstere{small-scale perpendicular gradients in} Fil-2 \commentstere{are revealed when calculating $\nabla$\vlsr  at scales of 2.5 times the beam} (Fig. \ref{fig:HC3N-nablav}). This type of velocity gradient is consistent with the contraction of a sheet-like cloud, as argued in filament simulations \citep{Chen2020filamentform-simulations}, \commentstere{as well as rotation around the filaments's main axis.} This \commentstere{type of} gradient has been observed in other filaments, for example, within Perseus and Serpens molecular clouds \citep{ FernandezLopez2014CLASSy-serpensfilaments,Dhabal2018filkinematics}. \commentstere{We interpret the velocity gradient as infall dominated gas because these filaments are not supported against gravitational collapse given the turbulence in the B5 core, unless a magnetic field is present, and show signs of contraction \citepalias{Schmiedeke2021supercriticalfil}. Infall of material toward the center of the filaments has been suggested for B5 in previous works which analyze the kinematics of} \ce{NH3}\,(1,1) \citep[][Choudhury et al. \commentstere{submitted}]{ Schmiedeke2021supercriticalfil, Chen2022B5_NH3_vel}. Our results \commentstere{support the interpretation that} the additional components found by Choudhury et al. \commentstere{(submitted)} are, as suggested, \commentstere{signs of} infall from the coherent core.   
\commentstere{We suggest that \ce{HC3N} is more sensitive to the flow of mass from the Barnard 5 core to the filaments, whereas \ce{NH3}, as it is a later type molecule, only traces the flow of gas within the filaments toward the condensations} and differentiates regions of subsonic and supersonic turbulent motions \citep{Pineda2010coherenceB5}. 

\subsection{A streamer toward B5-IRS1}

The two main \ce{H2CO} emission clusters in \refereetwo{position-position-velocity} space found in Sect. \ref{sec:dbscan-clustering} have velocity gradients consistent with infall motion. We confirmed that most of the emission on these structures is not affected by the outflow, at least beyond $\sim600$ au from the protostar. The nonthermal dispersion decreases as the distance to the protostar increases for both \refereecomment{clusters} (Fig.~\ref{fig:sigmantkde}). Most of the spectra located at distances $> 600$ au have subsonic $\sigma_{\mathrm{v,nt}}$. There are, however, some spectra with trans-sonic velocity dispersion ($\mathcal{M}_s \gtrsim 1$) within 600 au of the protostar. This result suggests that \ce{H2CO} gas located closer to the protostar is becoming affected by the outflow, which increases $\sigma_{\mathrm{nt}}$. 

We were able to confirm the infall nature of the blueshifted streamer using the streamline model. The best fit solution starts from the outer edge of the blue cluster's emission with null initial radial velocity. We note that the blue streamer is not a thin, long structure such as other streamers in the literature, but more similar to a bulk of gas infalling due to gravity. Moreover, as the channels that trace the streamer at a distance of about 2800 au from the protostar are affected by missing short-spacing data (as seen in Fig.~\ref{fig:H2CO-chanmap}, between 9.6 and 10.4 \kms), it is possible we miss some of the extent of this bulk. The spatial width of the emission we do trace suggests that the component could be fitted by a family of streamlines. We only took the one that best matches the general velocity KDE that passes through the emission seeen in the plane of the sky, because this allowed us to confirm its infalling nature. Nevertheless, this streamer has a similar length ($\approx2800$ au) to other streamers found toward Class I protostars \citep{Yen2014L1489Infall,Valdivia-Mena2022per50stream}. 

Understanding the kinematic nature of envelope emission helps disentangle the chemical history of the protostar. \commentstere{A} previous work by \cite{vantHoff2022snowline} shows \ce{H^{13}CO+} (2 -- 1) emission \commentstere{with} a ridge-like structure toward the east of B5-IRS1, similar to the arc shaped blueshifted envelope \refereecomment{cluster}. They suggest that emission is either associated with an extended water snowline due to a previous accretion burst in this source, or the larger protostellar envelope. We compare the \ce{H2CO} blueshifted envelope component with the \ce{H^{13}CO+} emission from \cite{vantHoff2022snowline} in Appendix \ref{ap:comparisonH13COp}. Both emissions overlap in the image plane, although \ce{H^{13}CO+} is more extended than \ce{H2CO}. The velocity gradients are the same and there are no significant differences between the central velocities of both molecular emissions. Therefore, we confirm that \ce{H^{13}CO+} traces the envelope's infalling motion and not \commentstere{an} extended snowline due to an accretion burst in this source. This suggest that using the \ce{H^{13}CO+} emission without further kinematic information about the source can overestimate the water snowline distance. This example highlights the importance of the kinematic information toward a source to interpret its chemical history.

\refereecomment{Confirming the streamer nature of the redshifted cluster is not as clear as for the blueshifted cluster.} The red \refereecomment{cluster} could be described using the same initial distance and angular velocity as the blue \refereecomment{cluster streamline model}. However, this model assumes that the blue and red \refereecomment{clusters} are part of the same envelope \commentstere{(i.e. have the same $\Omega_0$, $i$ and $PA$)}, which cannot be confirmed with the available data. Also, it is possible that the red streamer is more affected by missing short-spacings of our interferometric data than the blue streamer. In projected distance (i.e. in the RA-DEC plane), the streamline model of the blue streamer measures approximately 12\arcsec, so if the red streamer emission extends as suggested by the streamline model, the MRS of our data ($\approx 6\arcsec$, Sect. \ref{sec:obs-alma}) does not allow us to detect its emission and we are only detecting the brightest part. 
Therefore, the classification of ``streamer" is left as tentative for the red component of the envelope, and ALMA ACA observations of this region are required to confirm its streamer nature. 
\refereecomment{If confirmed, B5-IRS1 would have twin streamers} instead of only one as in other protostellar sources. This is similar to the case of L1489 IRS \citep{Yen2014L1489Infall}, a Class I source with two streamers observed in \ce{C^{18}O} (2 -- 1), and of [BHB2007] 11 \citep{Alves2020doublestream}, a \refereetwo{Class I/II} source with two streamers detected with \ce{^{12}CO} (2 -- 1). Simulations show that when a protostar is still embedded in its natal core, there are several asymmetric channels from which matter is funneled \citep[e.g.,][]{Padoan2014infallsim,Kuffmeier2017sim-ppdisks, Kuffmeier2023rejuvenatinginfall}, but in most cases, only one streamer is observed \citep[e.g.,][]{Pineda2020Per2streamer, Garufi2021arXiv-accretionDGTauHLTau, Valdivia-Mena2022per50stream}. The fact that we do not see multiple streamers \commentstere{in other sources} does not mean that \commentstere{they} are not present, it is possible that their column density is too low to be detected. Further discoveries of streamers will help determine if dual streamers are a common occurrence or an exceptional phenomenon.

\subsection{Connection between large scale and small scale infall \label{sec:disc-connectingscales}}

\begin{figure*}[htbp]
        \centering
        \includegraphics[width=0.49\textwidth]{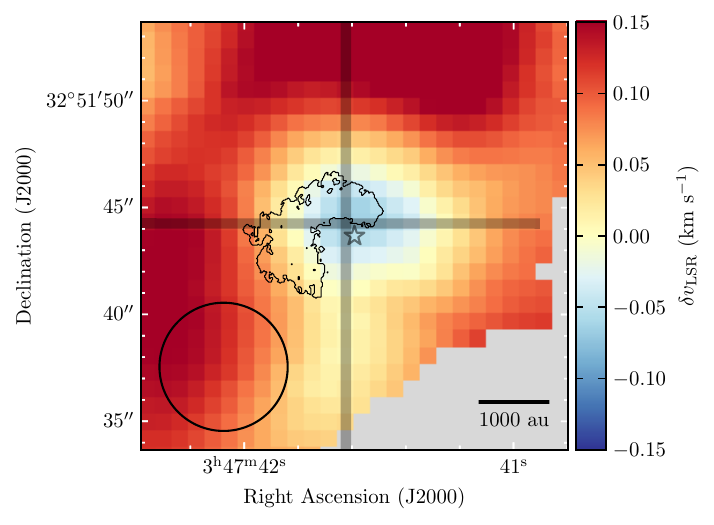}
        \includegraphics[width=0.49\textwidth]{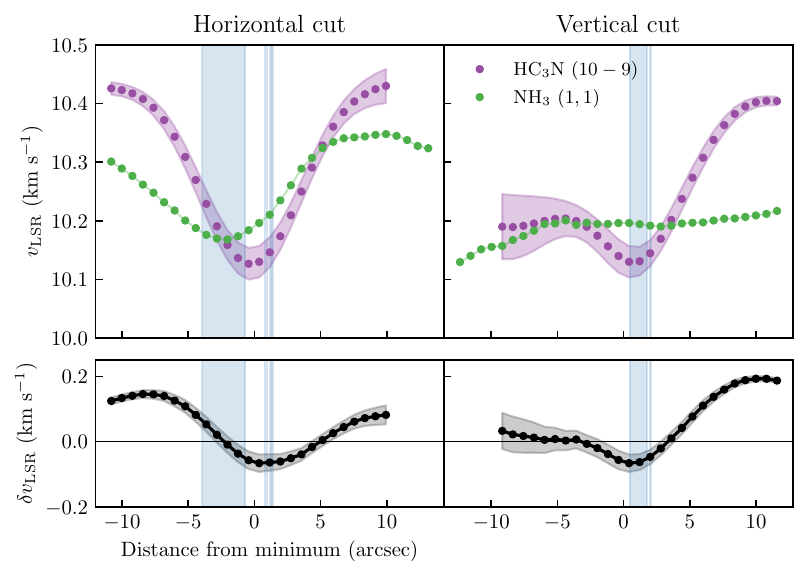}
        \caption{\textbf{Left:} Zoom into $\delta$\vlsr from Fig. \ref{fig:HC3N-NH3} toward the position of the protostar. The black contours mark the region of the blue streamer found in Sect. \ref{sec:dbscan-clustering}. The horizontal and vertical black lines show the positions of the horizontal and vertical cuts for the plots on the right. The black circle at the lower left corner represents the beam size of the \ce{NH3}\,(1,1) observations. \textbf{Right:} Horizontal and vertical cuts of \ce{HC3N} \vlsr and \ce{NH3} \vlsr. The purple points with the purple area represent the \ce{HC3N}\,($10-9$) \vlsr and its associated uncertainty after smoothing to reach the same resolution as the \ce{NH3}\,(1,1) \vlsr. The green points represent the \ce{NH3}\,(1,1) \vlsr (the uncertainties are roughly the size of the points). The black curves and black areas represent the difference between \ce{HC3N}\,($10-9$)   \ce{NH3}\,(1,1) \vlsr. The blue areas represent the location covered by the blue streamer.}
        \label{fig:HC3N-NH3-zoom}
\end{figure*}

In this work, we observed infall motions both at large scales, from the core to the filaments, and at small scales, from the protostellar envelope to the disk. \commentstere{Even if} the difference in resolution for these datasets is approximately tenfold, \commentstere{our results suggest that the confirmed streamer's origin lies within the fresh gas flowing from the B5 dense core}. We discuss the connection between these infall mechanisms.

Figure \ref{fig:HC3N-NH3} shows that the relative velocity of \ce{HC3N} with respect to \ce{NH3} reverses at the position of the protostar, showing a blueshifted zone about the size of the \ce{NH3} beam. Interestingly, zooming into this region (Fig. \ref{fig:HC3N-NH3-zoom} Left) reveals that this blueshifted area is located approximately at the position of the blue streamer found using \ce{H2CO} with ALMA data, which has a 10 times smaller \commentstere{beam}.  To \refereecomment{prove that the difference} between \ce{HC3N} and \ce{NH3} \refereecomment{bluehsifted emission} is significant, we show horizontal and vertical cuts of  \ce{HC3N}\,($10-9$) and \ce{NH3}\,(1,1) \vlsr, centered at the position of the minimum  $\delta$\vlsr, in Fig.~\ref{fig:HC3N-NH3-zoom}. The difference between both molecules' velocities is larger than the uncertainties of each fit: \ce{HC3N} decreases within a beam-sized area at the location of the protostar by approximately 0.4 \kms with respect to its surroundings, whereas \ce{NH3} has a decrease in \vlsr of about 0.1 \kms in the same region, only evident in the horizontal cut. Therefore,  \ce{HC3N}\,($10-9$) emission has larger variations around the protostar than \ce{NH3}\,(1,1). Moreover, this region coincides with where the AIC suggests two Gaussians fit \ce{HC3N} better than one Gaussian, but with not enough certainty (Fig. \ref{ap:fig-B5-zoom-prob}). The low certainty from the AIC criterion is \commentstere{caused by insufficient} spectral resolution to separate both components, but it suggests that, with better spectral resolution, we might be able to disentangle two velocity components, one related to the kinematics of the core gas infalling toward the filaments, and another, more blueshifted gas component which could trace the streamer. Overlaying the region where the streamer is present (blue vertical bands in Fig. \ref{fig:HC3N-NH3-zoom} \refereecomment{Right}), the streamer is close to the location of the minimum \vlsr for \ce{HC3N}. This coincidence suggest\commentstere{s} that \ce{HC3N} is sensitive to the infall motions \commentstere{close to the protostar} in the dense core. 

\commentstere{The velocity gradients seen in \ce{HC3N} also suggest the streamer is connected to chemically fresh gas.} The strongest  \ce{HC3N} velocity gradients are present \commentstere{toward Cond-1 and the protostar. A zoom into the north of Fil-2, shown in Fig.~\ref{fig:HC3N-gradient-zoom}, shows that gas flows along the filaments and reaches the condensations and the protostar, where they distribute between the three, as seen from the change in direction in the $\nabla$\vlsr vectors. The gradient orientation at the east of the protostar is similar to the orientation of the \refereetwo{velocity gradient of the blue cluster as seen in} Fig.~\ref{fig:dbscan-redblue-velocities}, \refereetwo{which in Sect. \ref{sec:slmodel} we determine is consistent with an accretion streamer.} Consequently, \ce{HC3N} could trace the streamer's motion.}

 \commentstere{These results suggest that the large scale flow traced by \ce{HC3N} connects to the tail of the streamer and, therefore, fresh gas coming from the B5 dense core reaches the protostellar disk through the streamer.} Previously discovered streamers have large extensions \citep[$\sim 10\,000$ au, e.g.,][]{Pineda2020Per2streamer} and for others, their full extension has not been imaged \citep{Valdivia-Mena2022per50stream}, showing that streamers can originate from great distances from the protostar. Due to the limited MRS of our data, this can be the case for B5-IRS1 as well.  \commentstere{However,} our current dataset's \commentstere{difference in resolution is too large to connect the two structures seamlessly}. \commentstere{Higher spectral resolution observations ($\lesssim 0.1$ \kms)} are required disentangle the second velocity component where the decrease in \vlsr occurs. Intermediate resolution ($\approx1\arcsec$) observations of \ce{HC3N}, together with and a spectral resolution comparable to our \ce{H2CO} data, \commentstere{could connect the flow from filament to streamer directly.}

\section{Summary\label{sec:summary}}

\begin{figure*}[htbp]
        \centering
        \includegraphics[width=0.99\textwidth]{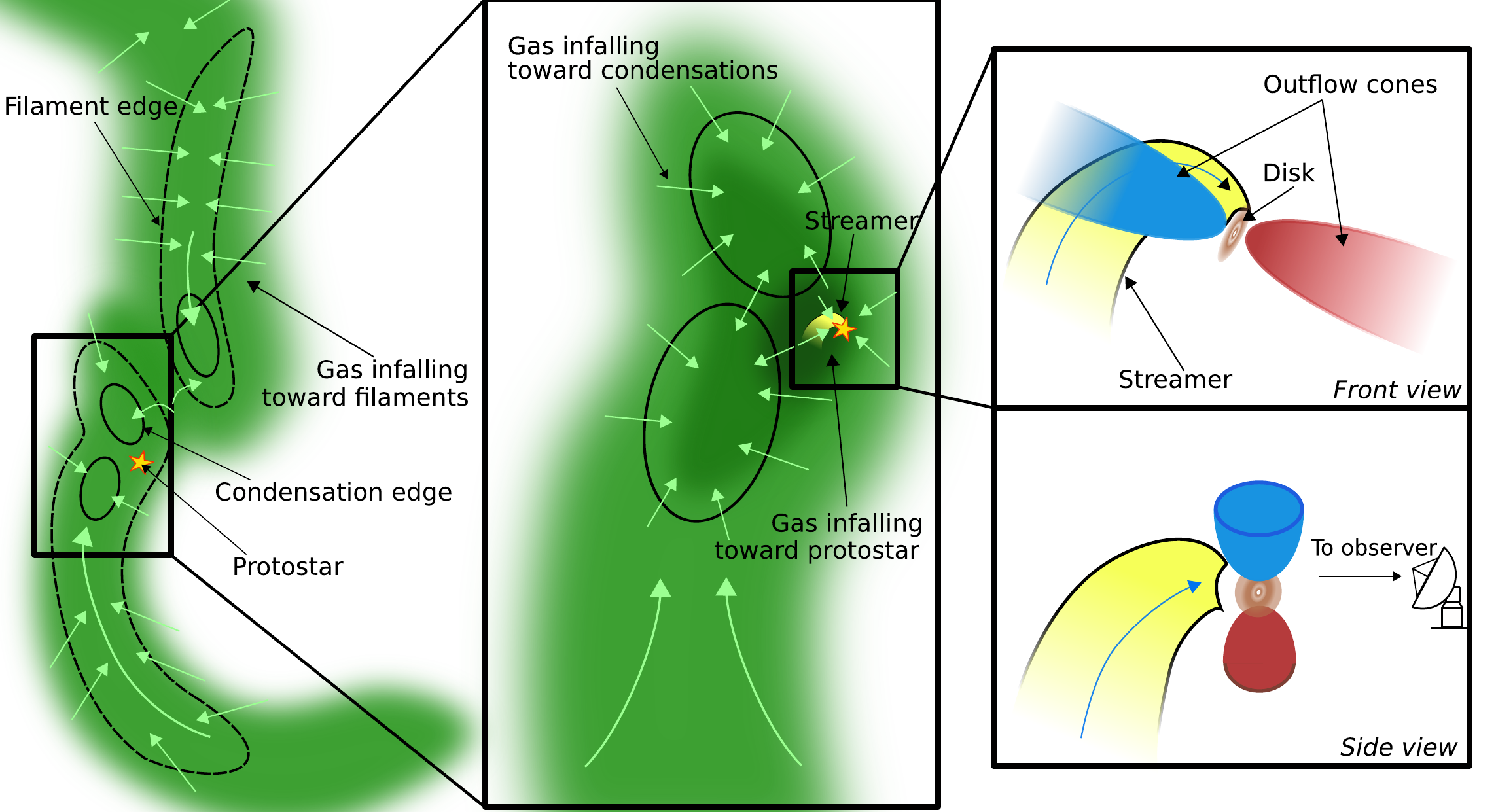}
        \caption{\commentstere{Diagram showing the flows of gas at the different scales investigated in this work. The yellow star indicates the position of the protostar. \textbf{Left:} Infall of fresh gas (green clouds) from B5 dense core toward the filaments (black dashed contours), with the flow direction indicated by the light green arrows. Black solid contours outline the condensations. \textbf{Center:} Infall of fresh gas from the filament toward the condensations (black solid contours) and the protostar, showing the change in direction from the filament to the condensations. The yellow curve shows the streamer transporting material toward the protostellar disk. \textbf{Top Right:} Infall of gas into the protostellar disk (brown disk) through the streamer (yellow flow), which is behind the blue outflow cone (blue cone). The blue arrow indicates the flow direction of the blueshifted streamer.  \textbf{Bottom right:} Same as in the top right, but rotated by 90\degr \ to observe the streamer unobstructed by the outflow cone. }}
        \label{fig:summary-drawing}
\end{figure*}

We studied the kinematic properties of gas at 4\arcsec ($\sim 1200$ au) scales toward the B5 coherent core filaments using \ce{HC3N}\,($10-9$) and ($8-7$) transition line maps, and 0\farcs4 ($\sim 120$ au) scales toward the protostellar envelope surrounding B5-IRS1 using \ce{H2CO} (3$_{0,3}$ -- 2$_{0,2}$) and \ce{C^{18}O} ($2-1$) line emission maps. We compared our results with previous \ce{NH3} emission observed toward this region. \commentstere{The main structures seen in \ce{HC3N} and \ce{H2CO} are summarized in Fig.~\ref{fig:summary-drawing}}. Our main results are summarized below.

\begin{enumerate}
    \item \textbf{\ce{HC3N} emission traces the gas flow from the coherent core to the filaments.} It is consistently redshifted with respect to the dense gas, observed in \ce{NH3}, which indicates that it follows different kinematic properties than the dense core tracer. \ce{HC3N}\,($10-9$) shows velocity gradients perpendicular to the filament length, which is consistent with the velocity profile expected from the contraction of a sheet-like cloud forming a filament \citep{Chen2020filamentform-simulations}.
    \item \refereecomment{Using \ce{HC3N}\,($10-9$) and ($8-7$) transitions, we obtain a mean H$_2$ volume density $n(\mathrm{H}_2)\approx10^{5}$ cm$^{-3}$. With this value, we estimate that \ce{HC3N} traces accretion rates of $\dot{m} = 4.3\times10^{-5}$ \Msun yr$^{-1}$ in Fil-1 and $\dot{m} = 3.5\times10^{-5}$ \Msun yr$^{-1}$ in Fil-2, similar to the accretion rates obtained in previous works which used \ce{NH3} emission.}

    \item \textbf{We find one streamer and one streamer candidate toward B5-IRS1, using a clustering algorithm on the velocity components found in \ce{H2CO} emission.} We confirm that \ce{H2CO} emission is mostly unaffected by the outflow cone in this region. We confirm the infalling nature of the clustered component which is blueshifted with respect to the protostar using a streamline model. The categorization of the component that is redshifted with respect to IRS1 is left as tentative as we only traced a small part of this infall. We estimate \refereecomment{that the} streamer has a total length of around 2800 au through the resulting parameters of the applied streamline model. 
    \item \textbf{We sugest that the infall of chemically fresh gas toward the condensations and filaments is connected to the protostar via the streamer.} At the location of the protostar, \ce{HC3N}\,($10-9$) central velocities decrease with respect to its surroundings. The location of the blueshift coincides with the location of the streamer \commentstere{and the velocity gradients seen in \ce{HC3N} coincide in orientation with the velocity gradient shown by the streamer}. We suggest \ce{HC3N} is also \commentstere{sensitive} to the infall at small scales, which we know are present because of the high-resolution observations. 
\end{enumerate}

Our results highlight the importance of the environment in the comprehension of the physical and chemical processes around protostars. The properties of both the core and the protostellar disk can be affected by the infall mechanisms we observe: there is both chemically fresh gas being deposited toward the filaments, as well as a preferential channel to deposit gas from the envelope to the disk. Intermediate spatial resolution observations are required to \commentstere{confirm} the large-scale infall traced using \ce{HC3N}\,($10-9$) with the small-scale infall seen in \ce{H2CO} emission.

\begin{acknowledgements}
        \refereecomment{The authors would like to thank the anonymous reviewer for their \refereetwo{careful review} of the paper and their constructive comments.}
          M.T.V., J.E.P. P.C. and S.C. acknowledge the support by the Max Planck Society. 
          D. S.-C. is supported by an NSF Astronomy and Astrophysics Postdoctoral Fellowship under award AST-2102405. 
          G.A.F acknowledges support from the University of Cologne and its Global Faculty programme. 
          S.S.R.O. acknowledges support from NSF Career award 1748571,  NSF AAG 2107942, NSF AAG 2107340, and  NASA grant 80NSSC23K0476.
          
          This work is based on observations carried out under project number S18AL with the IRAM NOEMA Interferometer and project 034-19 with the 30m telescope. IRAM is supported by INSU/CNRS (France), MPG (Germany) and IGN (Spain).
      This paper makes use of the following ALMA data: ADS/JAO.ALMA 2017.1.01078.S. ALMA is a partnership of ESO (representing its member states), NSF (USA) and NINS (Japan), together with NRC (Canada), MOST and ASIAA (Taiwan), and KASI (Republic of Korea), in cooperation with the Republic of Chile. The Joint ALMA Observatory is operated by ESO, AUI/NRAO and NAOJ. The National Radio Astronomy Observatory is a facility of the National Science Foundation operated under cooperative agreement by Associated Universities, Inc. This work made use of Astropy:\footnote{http://www.astropy.org} a community-developed core Python package and an ecosystem of tools and resources for astronomy \citep{astropy:2022}.
      
\end{acknowledgements}

%
\bibliographystyle{aa} 
\bibliography{main} 

%

\begin{appendix}
        
\section{Very Large Array and Green Bank Telescope\label{sec:obs-vla+gbt} observations}
        
\commentstere{We used the \ce{NH3}\,(1,1) inversion transition data from \citetalias{Pineda2015NatureB5cores} to define the filaments and cores and compare them to our NOEMA and 30m observations. The details of the reduction can be found in \citetalias{Pineda2015NatureB5cores} and \citetalias{Schmiedeke2021supercriticalfil}. In summary, VLA observations using the K-band were taken using D and CnD configuration (project number 11B-101), and GBT observations were included to recover the zero-spacing information in the UV plane (project number 08C-088). The data were reduced and imaged using CASA, with multiscale CLEAN and a Briggs weight (robust parameter of 0.5).}

\begin{figure}[htbp]
        \centering
        \includegraphics[width=0.49\textwidth]{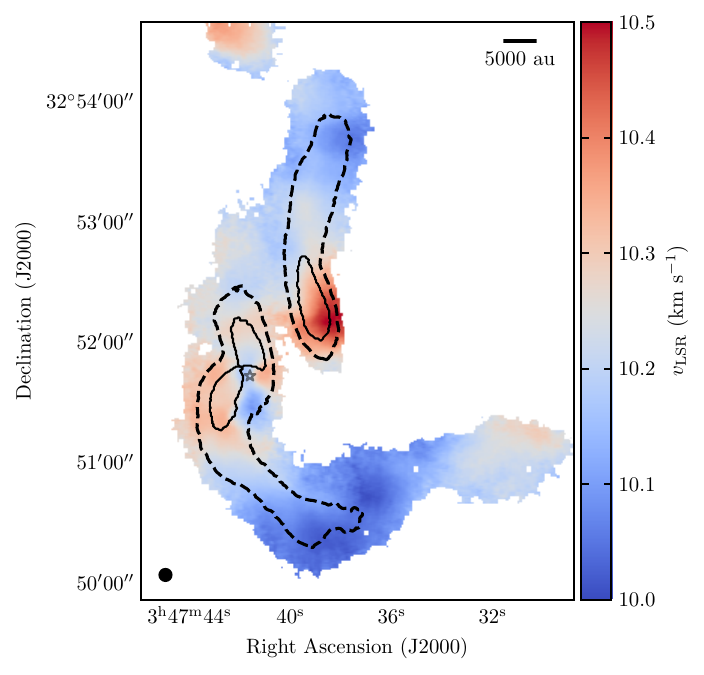}
        \caption{\commentstere{Central velocities of the \ce{NH3}\,(1,1) spectra from \citetalias{Pineda2015NatureB5cores}. The dashed lines outline the filaments seen in \ce{NH3}\,(1,1). The solid lines outline the condensations seen through the same molecule emission. The star marks the position of B5-IRS1. }}
        \label{fig:NH3-vlsr}
\end{figure}
        
\commentstere{Figure \ref{fig:HC3N-integrated} shows the contours of the dense filaments and condensations found in \ce{NH3}\,(1,1) emission from \citetalias{Pineda2015NatureB5cores} as defined in \citetalias{Schmiedeke2021supercriticalfil}, which used dendogram analysis to define the filaments and condensations. We use their nomenclature to name the filaments and cores throughout this work. Figure \ref{fig:NH3-vlsr} shows the central velocities \vlsr obtained in \citetalias{Pineda2015NatureB5cores} from a line fit to the \ce{NH3}\,(1,1) hyperfine components using the \texttt{cold-ammonia} model implemented in \texttt{pyspeckit} \citep{Ginsburg-Mirocha2011pyspeckit, Ginsburg2022pyspeckitpaper}. \refereecomment{The central velocities range from 10 to approximately 10.5 \kms and vary smoothly across the map. The central velocities' uncertainties are on average $7\times10^{-5}$ \kms and range from $3\times10^{-5}$ \kms within the condensations and $2\times10^{-4}$ \kms at the edges of the filaments.} Fil-1 shows a velocity gradient from 10.1 to 10.5 \kms approximately from north to south, and Fil-2, from 10 to 10.3 \kms from south to north, except near the protostar.}

\section{\refereecomment{\ce{H2CO} moment maps}\label{ap:H2COmoments}}

\begin{figure*}[htbp]
        \centering
        \includegraphics[width=0.49\textwidth]{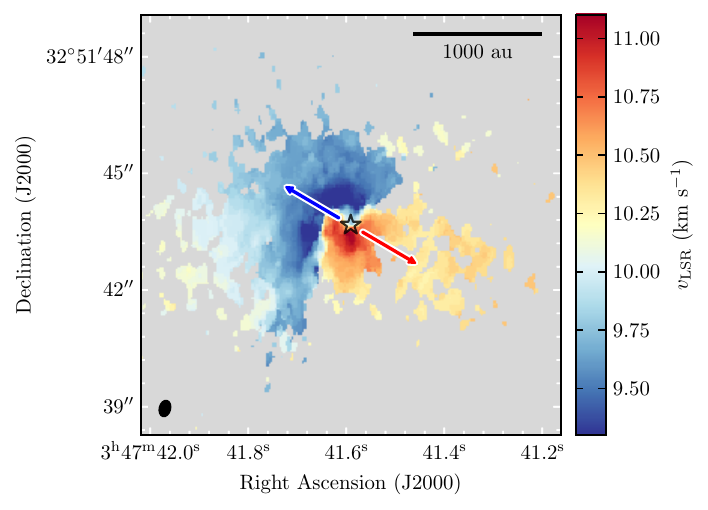}
        \includegraphics[width=0.49\textwidth]{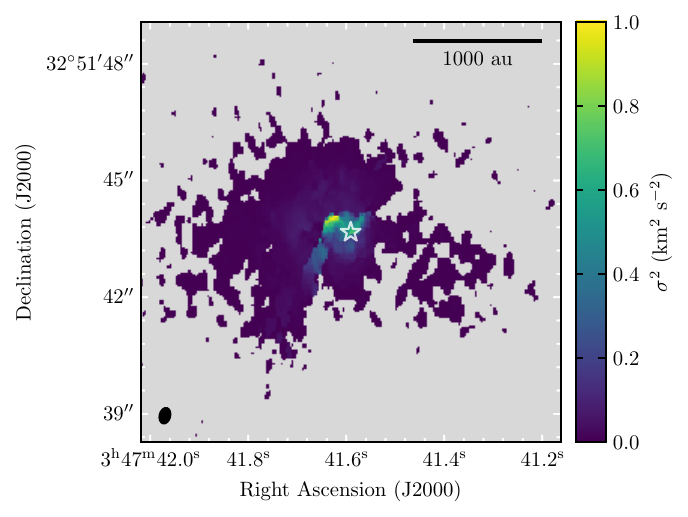}
        \caption{\refereecomment{Moments 1 (left) and 2 (right) of the ALMA \ce{H2CO} ($3_{0,3}-2_{0,2}$) line emission. The black (left) and white (right) stars represent the position of the protostar. The blue and red arrows indicate the directions of the blue and redshifted outflow lobes from \cite{Zapata2014B5outflow}. The scalebars indicate a length of 1000 au. The black ellipse in the bottom left corner represents the beam size. }}
        \label{fig:H2CO-momentmap}
\end{figure*}

\refereecomment{Figure \ref{fig:H2CO-momentmap} shows moments 1 (weighted velocity) and 2 (weighted velocity dispersion) for  \ce{H2CO} emission with $S/N>5$. The moment maps show that emission \refereetwo{has} different velocities in the east-west direction. Emmision toward the east of the protostar has a larger extension and is blueshifted with respect to B5-IRS1's \vlsr \citepalias[10.2 \kms,][]{Pineda2015NatureB5cores}. Emission toward the west covers less area and is mostly redshifted with respect to the protostar. The moment 1 map shows a curved boundary within a beam of the protostar for the blue and redshifted sides.}

\refereecomment{The moment 2 map shows that most \ce{H_2CO} emission has a variance $\sigma^2$ of about 0.1 \kms, except within a radius of approximately 0\farcs5 from the protostar and toward the south of it as well. Most notably, the peak of $\sigma^2$ is within a resolved distance (one beam) of B5-IRS1, toward the east. }

\section{\ce{C^{18}O} line emission images\label{ap:C18Odata}}

\begin{figure}[htbp]
        \centering
        \includegraphics[width=0.49\textwidth]{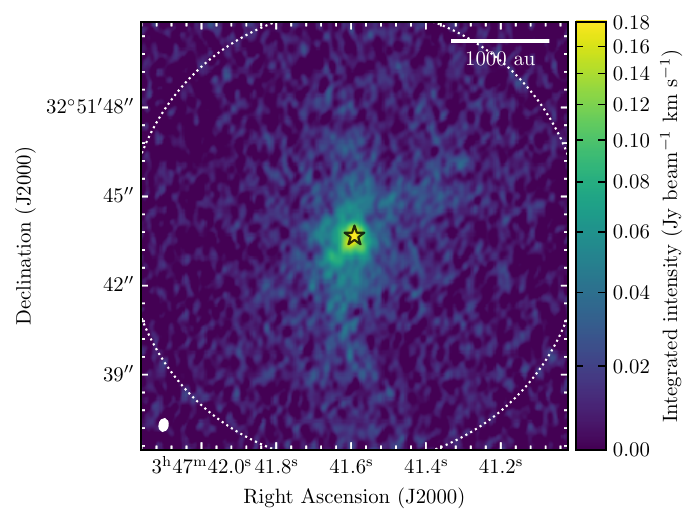}
        \caption{Velocity integrated image of \ce{C^{18}O} emission from \refereecomment{7.8 to 12.4} \kms. \commentstere{We note that the area covered in the \ce{C^{18}O} map is larger than for the \ce{H2CO} image, shown in Fig.~\ref{fig:H2CO-integrated}}. The black star represents the position of the protostar. The scalebar indicates a length of 1000 au. The white ellipse in the bottom left corner represents the beam size. The primary beam of the image is drawn with \commentstere{a white dotted circle.}}
        \label{fig:C18O-integrated}
\end{figure}

\begin{figure*}[htbp]
    \centering
    \includegraphics[width=0.95\textwidth]{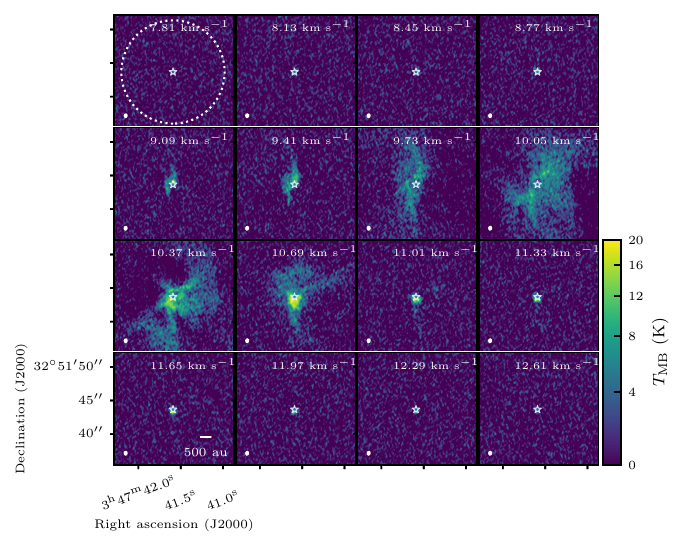}
    \caption{Channel maps between 7.81 and 12.61 \kms for the ALMA \ce{C^{18}O} (2 -- 1) spectral cube. The white star represents the location of the protostar. The white dotted circle marks the extent of the primary beam. The white ellipse at the bottom left corners represents the beam size. The scalebar shows a 500 au length in the map.}
    \label{fig:C18O-chanmap}
\end{figure*}

Figure \ref{fig:C18O-integrated} shows the integrated emission of  \ce{C^{18}O} from 7.8 to 12.4 \kms. Figure \ref{fig:C18O-chanmap} shows the channel maps of \ce{C^{18}O} between 7.8 and 12.3 \kms in 0.3 \kms steps.  \ce{C^{18}O} traces both the envelope and the natal cloud and is more extended than \ce{H2CO}.  Channels between 9.6 and 10.4 \kms are affected by strong bowls of negative emission due to missing short-spacing data, but \ce{C^{18}O} is more affected by the missing scales than \ce{H2CO}, as it has larger regions with negative emission artifacts. \ce{C^{18}O} shows point-like emission between 7.8 and 9.0 \kms and from 11.1 to 12.3 \kms, \refereecomment{which we suggest comes from the gas disk surrounding the protostar.}

\section{Gaussian fit to the spectra and selection criterion\label{ap:gaussfit}}

We fit a single Gaussian component to the spectra in the \ce{HC3N} (10--9) cube and one, two, and three Gaussian components in the H$_2$CO (3$_{0,3}$--2$_{0,2}$) cube, using the Python \texttt{pyspeckit} library \citep{Ginsburg-Mirocha2011pyspeckit}. We leave out of the analysis all spectra with a peak S/N lower than 5 for the case of \ce{HC3N} and lower than 3 for \ce{H2CO}.

We created a mask with the pixels to fit. First, we select all pixels with $S/N>5$. Then, we did a series of morphological operations on the mask using the \texttt{morphology} library from \texttt{scikit-image} python package \citep{scikit-image}. 
For \ce{HC3N} we did the following operations in order: first, we removed islands with fewer than 100 pixels using the function \texttt{remove\_small\_objects}. Secondly, we filled the holes smaller than 100 pixels with the function \texttt{remove\_small\_holes}. Finally, we did a morphological closing of the mask to fill cracks in the mask, using the \texttt{closing} function with a circular footprint with radius of 6 pixels.

For \ce{H2CO}, we built the mask with the following sequence: first, we removed islands with fewer than 50 pixels using the function \texttt{remove\_small\_objects}. Secondly, we filled the holes smaller than 50 pixels with the function \texttt{remove\_small\_holes}. Then, we did a morphological closing of the mask to soften cracks in the edges, using the \texttt{closing} function with a circular footprint with radius of 3 pixels. Afterwards, we filled the holes with fewer than 200 pixels. Finally, we did a morphological opening of the mask to round the edges using a disk footprint of radius 4 pixels. This mask allowed us to capture pixels with significant \ce{H2CO} emission, larger than $S/N>3$, that are connected to significant areas of emission and not in disconnected islands that are associated to artifacts.

The initial guesses are fundamental in all of our fits to obtain reasonable results. For \ce{HC3N}, we first obtained the moments of the cube, as described in the Minimal Gaussian cube fitting example from pyspeckit\footnote{https://pyspeckit.readthedocs.io/en/latest/example\_minimal\_cube.html}. Then, we checked the initial guesses to replace any value that is not within \refereecomment{acceptable ranges, as follows}. If the initial peak is negative, we replace it with a value of 0 K. If the initial central value was not within the range of observed velocities (between 9 and 11.3 \kms), we replaced it with 9 \kms if the value was lower than 9 \kms and with 11.3 \kms if the value was higher than 11.3 \kms. We replaced all initial velocity dispersions larger than 2.3 \kms to 2.3 \kms, which is the difference between the observed velocity boundaries. In the pixels where the initial guesses were not a number (NaN), we replaced the peak intensity for 1 K, the central velocity for 10.2 \kms and the dispersion for 1 \kms. 

The initial guesses for \ce{H2CO} fit were different in the cases we fit one, two and three Gaussians. For one Gaussian, we used the same initial guesses for all pixels: a peak temperature $T_{\mathrm{peak}}$ of 10 K, central velocity \vlsr of 10.2 \kms and dispersion $\sigma_{\mathrm{v}}$ of 0.8 \kms. For two Gaussians, we use the same initial guesses for most of the pixels except for select regions where we input different guesses manually. For most of the pixels, the first component had initial guesses of $T_{\mathrm{peak}}=10$ K, $v_{\mathrm{LSR}}=10.5$ \kms and $\sigma_{\mathrm{v}}=0.3$ \kms, and the second, $T_{\mathrm{peak}}=10$ K, $v_{\mathrm{LSR}}=9.2$ \kms and $\sigma_{\mathrm{v}}=0.3$ \kms. For three Gaussians, our initial guesses were $T_{\mathrm{peak}}=10$ K, $v_{\mathrm{LSR}}=11.3$ \kms and $\sigma_{\mathrm{v}}=0.3$ \kms for the first component, $T_{\mathrm{peak}}=10$ K, $v_{\mathrm{LSR}}=10.7$ \kms and $\sigma_{\mathrm{v}}=0.3$ \kms for the second, and $T_{\mathrm{peak}}=5$ K, $v_{\mathrm{LSR}}=9.5$ \kms and $\sigma_{\mathrm{v}}=0.3$ \kms for the third, except for selected regions where we input different guesses manually. 

After fitting once, we did a second fit, using the resulting output from the first fit as initial guesses. To do this, we selected for further analysis the fitted spectra that met all of the following requirements \refereetwo{and the rest are replaced with NaNs}: the parameter uncertainties had to be all smaller than 50\%; the amplitude of the Gaussian was a positive value; the Gaussian component had a central velocity in the observed emission velocity range (between 8 and 12 \kms for H$_2$CO, and between 9 and 11.3 \kms for \ce{HC3N}); and finally,  the fitted amplitude had $S/N>2$ for the case of \ce{H2CO} and $S/N>5$ for \ce{HC3N}.
Then, we interpolated the parameters obtained that passed this filter to fill in the pixels where the fit did not converge to a solution with the initial guesses, or where the first fit did not pass the filter described above. For the interpolation we used the \texttt{griddata} function from the \texttt{scipy.interpolate} python library \citep{2020SciPy-NMeth}. Afterwards, we input the interpolated values as initial guesses for the fits.

To mitigate the effects of the missing short- and zero-spacing information in our interferometric data for the \ce{H2CO} fit, we fit one, two and three Gaussian components using the spectra with the values between 9.7 and 10.3 \kms masked, following the same procedure above. \refereecomment{This only has a \refereetwo{positive} net effect on the final results for the spectra within a 500 au radius from the protostar.} For the rest of the spectra, masking the channels has no net positive effect on the final results and misses Gaussian components that peak within the masked channels. Therefore, we only used the fit with masked central channels for the spectra within a 500 au radius.

We kept the second fit results and did a further quality assessment. For \ce{HC3N}, the assessment included the requirements mentioned above. For \ce{H2CO}, we added the following criteria after evaluating the previous requirements: first, the fitted amplitudes selected all had $S/N>3$;  the uncertainty in the central velocity for each fit had to be less than the channel size (0.08 \kms for the \ce{H2CO} ALMA cube, Table \ref{tab:obsprops}); and finally, for pixels where the fit was done with masked channels, there could not be a peak located at the masked channels. After this evaluation, we also filtered islands of pixels in the image space that had fewer than 50 pixels for the one Gaussian fit, and fewer than 15 pixels for the two Gaussian fitting, as this would only add noise to the clustering. 

We used the Akaike information criterion (AIC) to decide how many Gaussian components reproduced best each spectra, in a similar manner as applied in \cite{Choudhury2020} and \cite{Valdivia-Mena2022per50stream}, for the case where we fit up to 3 Gaussians in \ce{H2CO}. This criterion uses the AIC value $AIC$ to determine which model minimizes information loss:
\begin{equation}
    AIC = 2k + \chi^2 + C
,\end{equation}
where $k$ is the number of free parameters of the model, $\chi^2$ is the classical chi-squared statistic and C is a constant defined by the number of independent channels and the uncertainties \citep{Choudhury2020}. Because each Gaussian component has three free parameters, $k=3g$, where $g$ is the number of Gaussian components in each model. For $C$, we assumed that each channel in the spectra has a constant normal error, which corresponds to the rms of each cube. As we used the same data to test the three models, C is the same for all models and does not play a role in choosing the best, so we set $C=0$. 

We evaluated the probability that the model with the minimum information loss, which is the one with the minimum $AIC$, is a considerable improvement from the other models for each spectrum. The probability that model $i$ is as good as the model with minimum $AIC$ to minimize information loss is proportional to the differencce between the minimum AIC, $AIC_{min}$ and the AIC value of model i, $AIC_{i}$:

\begin{equation}
    P = \exp \Big(\frac{AIC_{min}-AIC_{i}}{2}\Big) \label{eq:prob-aic}
.\end{equation}

For each spectrum, the fit with the lowest AIC value is the preferred one. We also checked that the best fit has at least a 95\% probability of minimizing the information loss better than when compared with any other of the models applied: this translates to a value of $P<0.05$. For the case of \ce{H2CO}, there are very few spectra (around 0.1\%) where the probability is less than 95\% \refereecomment{for the best model}, so we leave the fit with the lowest AIC value.

\subsection{\ce{HC3N} two Gaussian fit\label{ap:HC3Nsecondfit}}

\begin{figure}[htbp]
    \centering
    \includegraphics[width=0.45\textwidth]{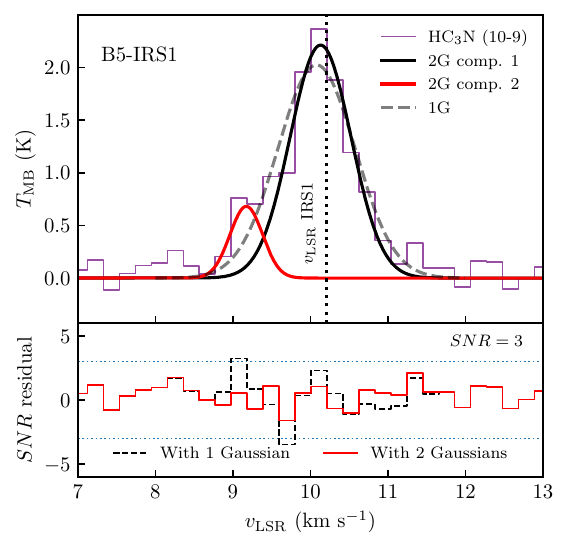}
    \caption{Results from the one- and two-Gaussian fit to the \ce{HC3N} (10--9) emission at the position of the protostar B5-IRS1. Top: \ce{HC3N} (10--9) emission spectrum at the position of B5-IRS1, together with the best fit results using one and two Gaussian curves.  The vertical dashed line represents the central velocity of the protostar from \citetalias{Pineda2015NatureB5cores}. Bottom: residuals from the fit in terms of $S/N$. The dashed purple line represents the residuals after the 1 Gaussian fit, whereas the solid purple line, the residuals after the 2 Gaussian fit. The horizontal blue dashed lines mark the $S/N=3$ and $-3$ levels.}
    \label{fig:ap-HC3N-protostarfit}
\end{figure}

\begin{figure}[htbp]
    \centering
    \includegraphics[width=0.45\textwidth]{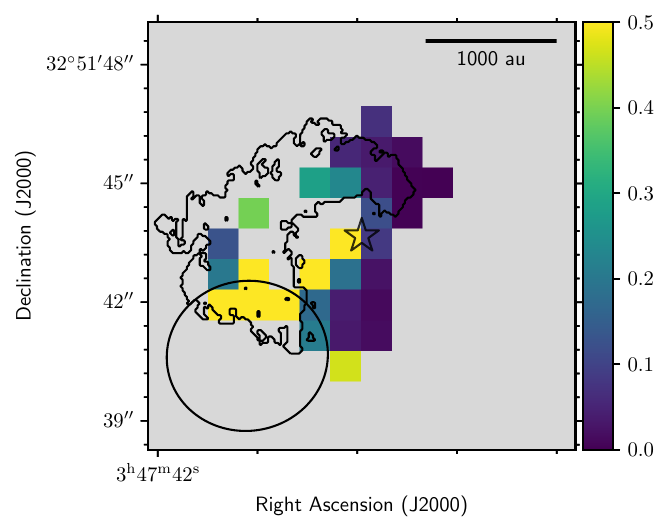}
    \caption{Probability $P$ (Eq.~\ref{eq:prob-aic}) for two Gaussian components fit in \ce{HC3N} (10--9) around 4\arcsec from the protostar. Contours indicate the location of the blue streamer found in Sect. \ref{sec:dbscan-clustering}.
    \label{ap:fig-B5-zoom-prob}}
\end{figure}

The best fit parameters for the \ce{HC3N} emission spectra are in Figure \ref{fig:HC3N-fitparams}. Figure \ref{fig:ap-HC3N-protostarfit} shows the spectrum at the position of the protostar and its Gaussian fit \refereecomment{with a dashed line}. We also did a two Gaussian fit to the central 4\arcsec around the protostar in the \ce{HC3N} map. This possible second feature is hinted at in the residuals of the one Gaussian fit at the position of the protostar, seen in Fig.~\ref{fig:ap-HC3N-protostarfit}, where some residuals reach $S/N\gtrsim3$. We fit two Gaussians and compare the results using the Akaike Information Criterion (AIC). The resulting fit at the location of the protostar is shown in Fig. \ref{fig:ap-HC3N-protostarfit} \refereecomment{with solid lines}. The residuals from both fits show that two Gaussians fit the spectrum better than one at the position of the protostar. However, as too few spectra were actually better fitted with two Gaussians that with one when evaluated with AIC (less than 10\% of the pixels, Fig. \ref{ap:fig-B5-zoom-prob}), we decided to leave all spectra with only one Gaussian.

\section{Core and filament \refereecomment{volume density} traced by \ce{HC3N} \label{sec:coremass} }

\refereecomment{We obtained the volume density of \ce{H2} gas $n(H_2)$ using the \ce{HC3N} line ratio between the 10 -- 9 and 8 -- 7 level transitions and the non-LTE radiative transfer code RADEX \citep{vanderTak2007RADEX}, with the goal of calculating the infall rate of material from the core to the filaments in Barnard 5. }

\subsection{\ce{HC3N} (8 -- 7) integrated image\label{sec:HC3N87line}}

\begin{figure}[htbp]
        \centering
        \includegraphics[width=0.49\textwidth]{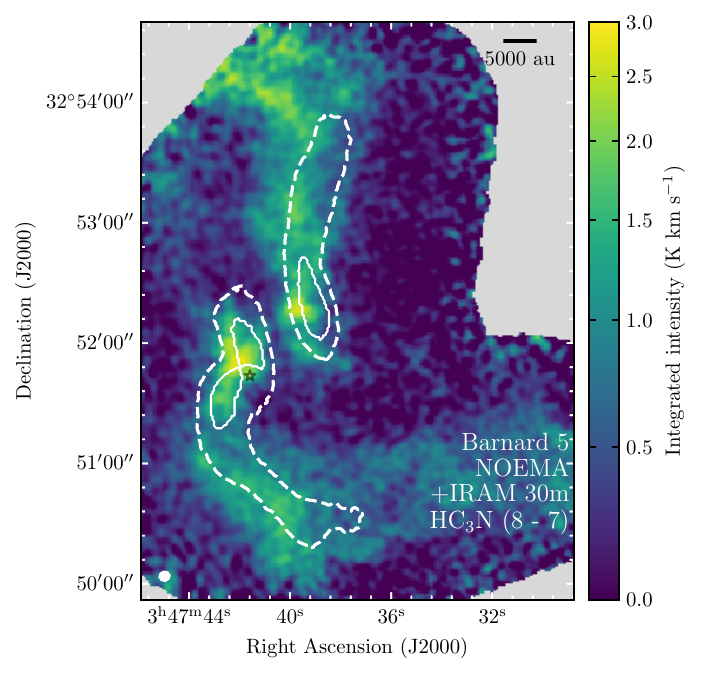}
        \caption{Velocity integrated \ce{HC3N} $(8-7)$ line emission, from 9.2 to 11.2 \kms. White dashed contours correspond to the filaments identified in \ce{NH3} emission by \citetalias{Pineda2015NatureB5cores} and \citetalias{Schmiedeke2021supercriticalfil}. White \commentstere{solid} contours \commentstere{outline} the edges of the condensations labeled as in \citetalias{Pineda2015NatureB5cores}. The \refereecomment{gray} \commentstere{star} marks the position of the protostar B5-IRS1.}
        \label{fig:HC3N-8-7}
\end{figure}

\refereecomment{Figure \ref{fig:HC3N-8-7} shows the integrated intensity of \ce{HC3N} $(8-7)$ emission. We integrated the line cube between 9.2 and 11.2 \kms, the same range as for \ce{HC3N} $(10-9)$. The morphology is similar to the \ce{HC3N} $(10-9)$ integrated map, but the peak in integrated intensity is located within Cond-2 instead of the protostar. }

\subsection{\ce{HC3N}  line ratio  \label{sec:HC3Nlineratio} }

\begin{figure}[htbp]
        \centering
        \includegraphics[width=0.49\textwidth]{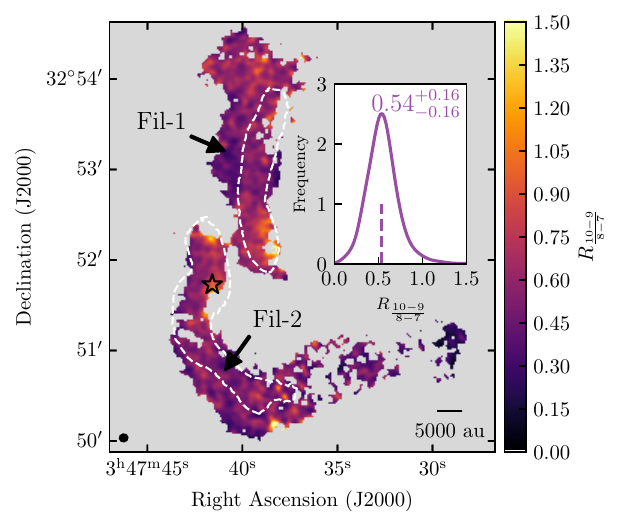}
        \caption{Map of the line ratio between the \ce{HC3N}\,($10 - 9$) and (8 -- 7) transitions. \refereecomment{White} dashed contours correspond to the filaments identified in \ce{NH3} emission by \citetalias{Pineda2015NatureB5cores} and \citetalias{Schmiedeke2021supercriticalfil}. The \refereecomment{black star} marks the position of the protostar B5-IRS1. Inset: KDE of the line ratios obtained. The dashed vertical line shows the median value of the distribution, and the uncertainties correspond to the distance to the 1st and 3rd quartiles. }
        \label{fig:HC3Nratio}
\end{figure}

We used the integrated line intensity maps for both  \ce{HC3N} transitions to obtain a line ratio map. First, \refereecomment{we smoothed} the \ce{HC3N}\,($10-9$) integrated image with a Gaussian kernel in CASA to the reach the resolution of the ($8-7$) transition map ($\approx5\arcsec$, Table \ref{tab:obsprops}). \refereecomment{Then, we} regridded the resulting convolved image to match the ($8-7$) line map pixel grid. Finally, we divided the regrid \ce{HC3N}\,($10-9$) integrated intensity map by the \ce{HC3N} ($8-7$) integrated image:

\begin{equation}
        R_{\frac{10-9}{8-7}} = \frac{ \int T_{R,10-9}dv }{\int T_{R,8-7}dv},
\end{equation}
where $\int T_{R}dv$ is the integrated intensity of a spectrum in K \kms.  The resulting map is in Fig. \ref{fig:HC3Nratio} and includes pixels where $S/N>5$ for both $10-9$ and $8-7$ transitions. 

In general, $R_{\frac{10-9}{8-7}}$ does not present large systematic variations along the filaments, and variations are usually clumpy and reach \refereecomment{differences of} up to 0.2 within one beam. The distribution of line ratios of the whole map is shown in the inset of Fig. \ref{fig:HC3Nratio}: it has a median of 0.54 and its shape resembles a Gaussian. 
The ratio does reach values up to 1.0 in the vicinity of the protostar, which could occur due to higher temperatures \refereecomment{and/or higher optical thickness} in the region. Very low values of $R_{\frac{10-9}{8-7}}$, under 0.2, and very high values, over 1.1, are located in the edges of the filaments and are due to the detection edges of either \ce{HC3N}\,($10-9$) or \ce{HC3N} ($8-7$). Near the condensations the ratio is \refereecomment{also} slightly higher, increasing from $\approx 0.55$ to $\approx 0.65$. This could be due to a higher temperature closer to the condensation, caused by mass infall, which would then populate more the ($10-9$) transition,  \refereecomment{and/or emission in the condensations becoming optically thick}. As the map shows no considerable spatial variations, it is safe to take its mean to describe the \refereecomment{volume density} of the whole emission.

\subsection{\refereecomment{\ce{H2} volume density traced by \ce{HC3N}} \label{ap:column-dens-details}}

We took the mean value of the distribution, $<R> = 0.56 \pm 0.01$, to estimate the H$_2$ density $n(H_2)$ and the excitation temperature $T_{ex}$ for the \ce{HC3N}\,($10-9$) transition. We \refereecomment{compared this ratio with} the results from the RADEX models run by \cite{Pineda2020Per2streamer}, as their input parameters result in similar emission intensities $T_{R}$ \refereecomment{as the ones observed in this work and the line ratios resulting from their test (between 0.2 and 0.8) include our resulting line ratio.} A line ratio $<R> = 0.56 \pm 0.01$ corresponds to a density of $\approx10^5$ cm$^{-3}$ (based on the Extended data fig. 1 from \cite{Pineda2020Per2streamer}) \refereecomment{and results in an excitation temperature $T_{ex}\approx6.4$ K for the $10-9$ molecular transition.}

\section{\refereecomment{Clustering of \ce{H2CO} Gaussian components\label{ap:clustering}}}

\begin{figure*}[htbp]
        \centering
        \includegraphics[width=0.33\textwidth]{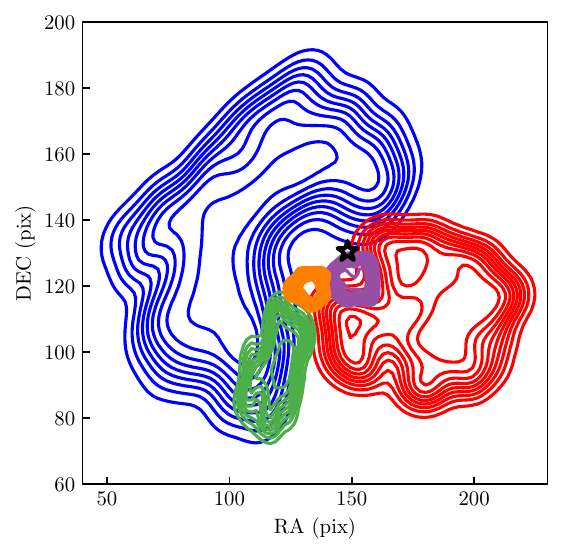}
        \includegraphics[width=0.33\textwidth]{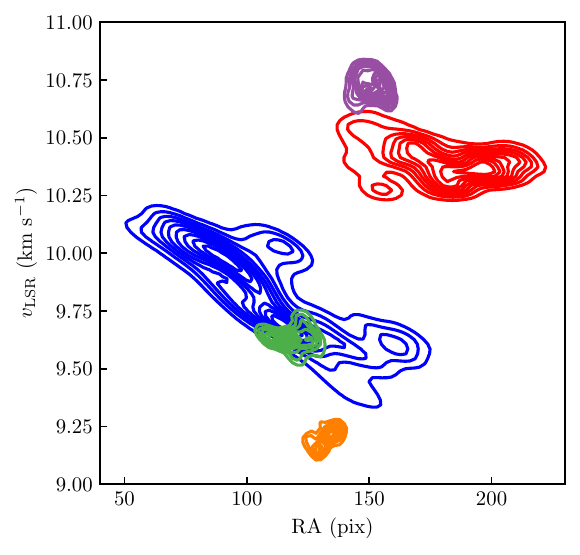}
        \includegraphics[width=0.33\textwidth]{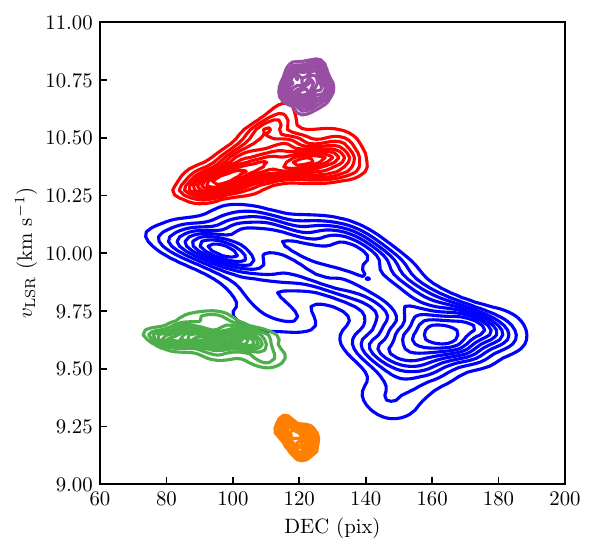}
        \includegraphics[width=0.97\textwidth]{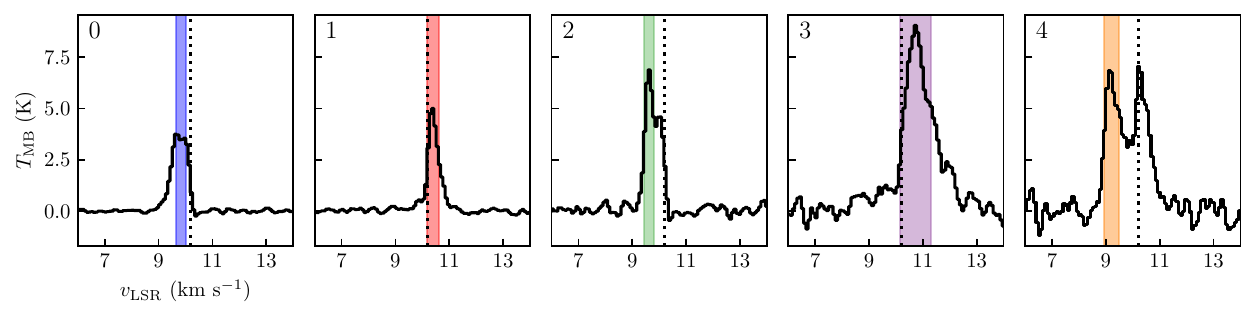}

        \caption{ \commentstere{Results from the clustering of the Gaussian components in the ALMA \ce{H2CO} spectra.} \commentstere{Each cluster is labeled with a number from 0 to 4, in order of descending number of points.}  Blue and red points (labeled 0 and 1) represent the clusters named blueshifted and redshifted component, respectively. Orange, purple and green points represent the other three found clusters. \textbf{Top}: \refereecomment{Density plots of the clustered groups using Gaussian KDEs, seen in the image plane (RA and DEC in pixels, left), in RA versus \vlsr (middle) and in DEC versus \vlsr (right).} \refereecomment{The black star marks the position of the protostar}. \commentstere{\textbf{Bottom:} average spectra \refereecomment{taken from the area covered by} each group, labeled with the numbers from the Top Left panel. The ALMA \ce{H2CO} spectra are drawn with black solid lines. The colored vertical areas represent the location of average \vlsr and FWHM of each cluster. The vertical dotted line marks the protostar's \vlsr. }}
        \label{fig:dbscan-total}
\end{figure*}

\refereecomment{We clustered the different Gaussian components found in Sect. \ref{sec:h2co-fitting} to aid in their physical interpretation. We applied the unsupervised learning tools from the \texttt{scikit-learn} Python library \citep{scikit-learn} to the results of our 1, 2 and 3 Gaussian component fitting of the ALMA \ce{H2CO} emission.}
We used the right ascension (RA), the declination (DEC) and central velocities (\vlsr) of each Gaussian component as features for clustering. Using all properties of the fits as features, or using the three mentioned parameters plus either amplitude or velocity dispersion, does not represent an improvement in the clustering of the \commentstere{Gaussian components}. We first scaled all the features to a range from -1 to 1 using the \texttt{StandardScaler} class in the \texttt{preprocessing} \commentstere{Python} library. By doing so, we assigned the same \refereecomment{relative} weight to all the features. Then, we used the algorithm DBSCAN \citep{Ester1996DBSCAN}, implemented in the \texttt{DBSCAN} class, with parameters \commentstere{\texttt{eps} $=0.248$} and \texttt{min\_samples} $=100$. The rest of the parameters are left as default. This results in 5 clusters in (RA, DEC, \vlsr) space. \commentstere{The resulting groups found by DBSCAN are} shown in Fig.~\ref{fig:dbscan-total}.

There are a total of \commentstere{11992} individual Gaussian components in the cube, out of which DBSCAN determined \commentstere{3042 (25\%)} are outliers, and do not belong to any cluster. \commentstere{This does not imply that these Gaussians are fitting noise, only that the features could not be grouped. In particular, points within 1\arcsec to the east of the protostar could not be grouped. \refereecomment{This region is dominated by emission with \vlsr$\approx 9.25$ \kms}. We suggest this emission is also part of the blueshifted streamer described in Sect. \ref{sec:slmodel}, as it follows the same apparent gradient. Adding these points to the blueshifted cluster does not affect the streamline model nor the resulting centrifugal radius, so we decided to follow the grouping resulting from the clustering algoithm.} \refereecomment{Out of all the clustered points, the largest group is a fully blueshifted cluster (consisting of \commentstere{5529} individual points) and the second largest, a fully redshifted cluster (with \commentstere{2557} individual points), with respect to the protostar's \vlsr(10.2 \kms). }

\subsection{Nonthermal velocity dispersion of \ce{H2CO} emission \label{ap:nonthermalH2CO}}

We checked the velocity dispersion $\sigma_{\mathrm{v, fit}}$ of the Gaussians in the blueshifted and redshifted \refereecomment{clusters} and compared them to the thermal sound speed. Nonthermal velocity dispersion larger than the sound speed can be attributed to the outflow or to gas tracing rotation motions instead of infall. 

First, we deconvolved the velocity dispersion with respect to the channel size of our data using
\begin{equation}
        \sigma_{\mathrm{v}}^2 = \sigma_{\mathrm{v,\,fit}}^2 - \frac{2\Delta_{\mathrm{ch}}}{2.355}
,\end{equation}
where $\Delta_{\mathrm{ch}}$ is the channel width (Table \ref{tab:obsprops}). Then, we subtracted in quadrature the thermal dispersion $\sigma_{\mathrm{th}}$ from the deconvolved velocity dispersion, thus obtaining $\sigma_{\mathrm{v,\,nt}}$:
\begin{equation}
        \sigma_{\mathrm{v,\,nt}}^2 = \sigma_{\mathrm{v}}^2 - \sigma_{\mathrm{th}}^2.
\end{equation}
The thermal dispersion is
\begin{equation}
        \sigma_{\mathrm{th}} = \sqrt{\frac{k_B T}{\mu_{\mathrm{H}_2\mathrm{CO}} m_\mathrm{H}}} = 0.052\, \mathrm{ km\,s}^{-1},
\end{equation}
where $k_B$ is the Boltzman constant, $T$ is the kinetic temperature of the gas, which we assumed is 9.7 K from \cite{Pineda2021B5ions}, $\mu_{\mathrm{H}_2\mathrm{CO}}=30.026$ is the molecular weight of the \ce{H2CO} molecule and $m_\mathrm{H}$ is the mass of a Hydrogen atom. 

\begin{figure}[htbp]
        \centering
        \includegraphics[width=0.49\textwidth]{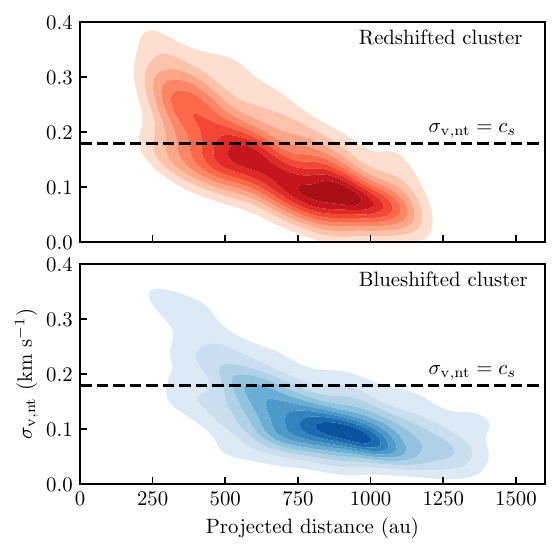}
        \caption{KDE of the nonthermal velocity dispersion for the blueshifted and redshifted envelopes. The dashed line represents where $\sigma_{\mathrm{v, \,nt}}$ equals the sound speed for gas at 9.7 K (0.18 \kms).}
        \label{fig:sigmantkde}
\end{figure}

The resulting $\sigma_{\mathrm{v,\,nt}}$ for the two envelope components are in Fig.~\ref{fig:sigmantkde}. The dispersion increases when the distance to the protostar decreases for both components, but it is mostly subsonic for both parts of the envelope. \ce{H2CO} emission becomes transonic (larger than $c_{\mathrm{s}}=0.18$ \kms) within $r \lesssim 600$ au approximately. This means that, for majority of the emission within both components, the \ce{H2CO} gas is unaffected by the outflow and is not caused by Keplerian rotation.

\section{Estimate of the protostellar mass \label{sec:starupperlim}}

\begin{figure}[htbp]
        \centering
        \includegraphics[width=0.49\textwidth]{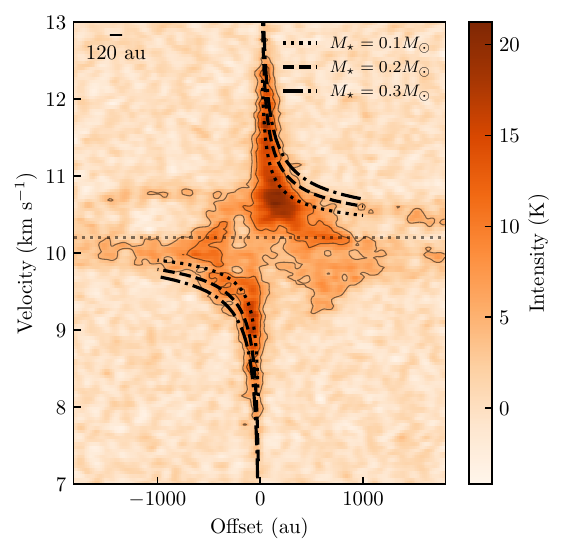}
        \caption{Position-velocity diagram of the ALMA \ce{C^{18}O} datacube, centered at the position of the protostar, with a PA of 157\fdg1 and a width of \refereetwo{0\farcs4}. The black curves show the Keplerian rotation profile produced by three protostellar massses: the \refereecomment{dotted} line shows the Keplerian curve for a protostar of 0.1 \Msun\citep{Brassfield2011B5envmass}, the \refereecomment{dashed} line for 0.2 \Msun, \refereecomment{and the dash-dotted line for 0.3 \Msun}. The dotted horizontal line marks the protostar's $v_{\mathrm{LSR}} = 10.2$ \kms. \refereecomment{The scalebar represents one beam width (120 au or 0\farcs4)}}
        \label{fig:c18o-pv}
\end{figure}

We used the \ce{C^{18}O} ($2-1$) ALMA data to estimate the protostellar mass of B5-IRS1. We first looked at the position-velocity (PV) diagram of \ce{C^{18}O} using the astropy package \texttt{pvextractor} \citep{pvextractor}. We generated the PV diagram along a path centered on the protostar with a total length of 1600 au, with a position angle (PA) of 157\fdg1 (with 0\degr\, pointing toward the north), obtained from the outflow PA \citep{Zapata2014B5outflow}, and a width of \refereetwo{0\farcs4}. 
\refereecomment{The pixel size of the \ce{C^{18}O} cube is 0\farcs054, so we chose to take an integer number of pixels as the width of the PV sampling path that was as close to the beam width as possible (\refereetwo{7} pixels, or \refereetwo{0\farcs4}).}
The resulting PV diagram is in Fig.~\ref{fig:c18o-pv} and shows the classical diamond shape that is attributed to a combination of Keplerian rotation and infall motion of the inner envelope \citep[e.g.,][]{Lee2014HH212rotinfall, Sakai2014Nature}. When we compare the PV diagram with the Keplerian curve produced by a 0.1 \Msun protostellar mass \citep[\refereetwo{the previously reported protostellar mass in}][]{Brassfield2011B5envmass} and for a disk with an inclination $i=77\degr$ \citep[obtained from the outflow inclination angle $i=13\degr$,][]{Yu1999B5outflow}, the PV diagram 5$\sigma$ limit is above the curve, indicating the protostar's mass is possibly higher. We plotted over the PV diagram the Keplerian curve for 0.2 \Msun with the same disk $i$ and we saw that it is closer to the 5$\sigma$ edge than 0.1 \Msun. From this simple observation, we suspect the central protostar's mass \refereetwo{is} be higher. To confirm, we did a deeper analysis into the \ce{C^{18}O} data.

We obtained the central position of \ce{C^{18}O} emission in each velocity channel and determine if it is consistent with Keplerian motion. For this, we fit a two-dimensional Gaussian at each of the channels of the \ce{C^{18}O} ($2 - 1$) cube between 7.9 and 9.2 \kms and between 10.8 and 12.4 \kms. These ranges cover the velocities of \ce{C^{18}O} emission that are the most blueshifted and redshifted with respect to the protostar, and miss the central channels seen in Fig.~ \ref{fig:C18O-chanmap}, where emission is affected by missing short- and zero-spacings. We used the \texttt{models.Gaussian2D} class and fit the Gaussian model using least-squares minimization with the Levenberg-Marquardt algorithm, implemented in the \texttt{fitting.LevMarLSQFitter} class. Both classes are part of the \texttt{astropy} python package. We use the peak value of each channel, the position of the peak, and the beam width $\sigma_{\mathrm{beam}} = \theta_{\mathrm{FWHM}}/2.355$ (Table \ref{tab:obsprops}) as initial guesses for the amplitude, central position and dispersion of the Gaussian at each channel. We used the channels between $v_{\mathrm{LSR}} = 7.9 - 8.8$ \kms and $11.1 - 12.4$ \kms to calculate the barycenter (i.e. the rotation center) of the points blueshifted and redshifted with respect to the protostar's $v_{\mathrm{LSR}}$. Finally, we calculated the distance of each point to the calculated barycenter $d$ and plot $d$ against the corresponding channel's velocity with respect to the protostar's $v_{\mathrm{LSR}}$, $\delta v_{\mathrm{LSR, IRS1}} = v_{LSR} - 10.2$ \kms. Using the corrected center instead of the position from \cite{Tobin2016VANDAMmultiples} returns a more precise velocity versus distance plot. Nevertheless, the distance between the barycenter and the protostar's coordinates is $\lessapprox 0\farcs1$.

\begin{figure*}[htbp]
        \centering
        \includegraphics[width=0.45\textwidth]{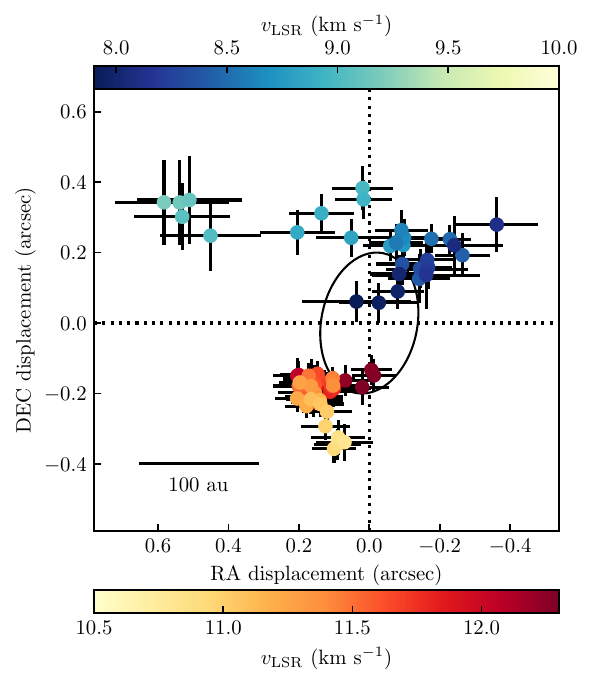}
        \includegraphics[width=0.54\textwidth]{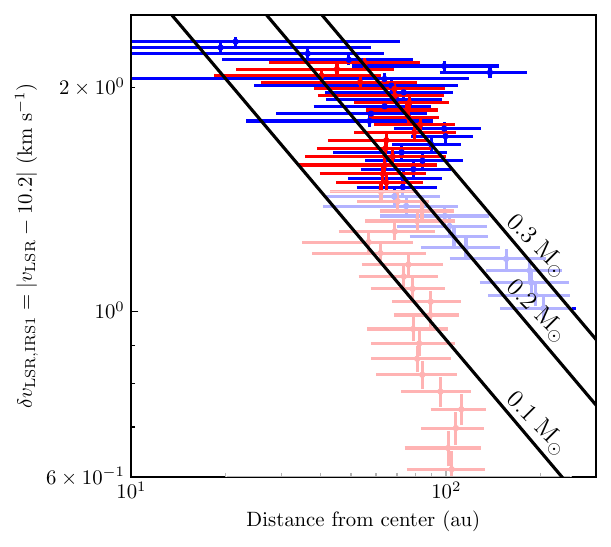}
        \caption{Results from the two-dimensional Gaussian fit to the \ce{C^{18}O} channel maps. \textbf{Left}: Best fit central positions with respect to the image plane, where their colors represent the velocity of the channel \vlsr. The dashed lines represent the position of the barycenter. The ellipse centered at the barycenter represents the beam size (FWHM). \textbf{Right}: \ce{C^{18}O} emission velocity with respect to the protostar's \vlsr versus distance from the protostar. Blue dots are obtained from the \ce{C^{18}O} emission blueshifted with respect to the protostar and red dots from the redshifted emission. Faded dots come from velocity channels with considerable extended emission, and therefore their properties are affected by motions other than Keplerian. The solid black lines are the Keplerian rotation profiles for \refereecomment{0.1,} 0.2 and 0.3 \Msun. }
        \label{fig:C18O-gaussfit}
\end{figure*}

The resulting \ce{C^{18}O} central positions per velocity channel are in Fig.~\ref{fig:C18O-gaussfit} Left and the velocity versus distance plot is in Fig.~\ref{fig:C18O-gaussfit} Right. The faded blue and red points represent the fitted positions where there is extended emission in the velocity channel, which affected the Gaussian fit. For emission peaks located from 0 to about 70 au (0\farcs2), as distance from the protostar increases, $\delta\mathrm{v}_{LSR, \mathrm{IRS1}}$ tends to decrease. In the image plane, as $\delta\mathrm{v}_{LSR, \mathrm{IRS1}}$ decreases, the points tend to get away from the protostar in opposite directions along a northwest to southeast orientation, similar to the orientation of the disk used to obtain the PV diagram. These two behaviors are expected for Keplerian rotation. However, emission centered farther away than $\sim70$ au show a change in direction: as \vlsr gets closer to the protostar's \vlsr, the blueshifted points move toward the east instead of northwest, and the redshifted points move toward the west instead of southeast. These low $\delta\mathrm{v}_{LSR, \mathrm{IRS1}}$ ``tails'' have different behaviors in velocity versus distance: $\delta\mathrm{v}_{LSR, \mathrm{IRS1}}$ drops faster with distance for the redshifted points than for the blueshifted ones. Moreover, the blueshifted and redshifted points that are the farthest away from the protostar are at the edges of the blueshifted and redshifted envelope components, respectively. We note that these points come from the channels which have considerable extended emission. This suggest that \ce{C^{18}O} emission is tracing part of the infall that we trace with \ce{H2CO} emission as \ce{C^{18}O} extended emission.

We plotted the Keplerian velocity versus distance curve expected for a protostellar mass of 0.1, 0.2 and 0.3 \Msun over the 
velocity versus distance plot for \ce{C^{18}O} positions (in Fig.~\ref{fig:C18O-gaussfit} Right). \refereecomment{The uncertainty in central position is too large to fit a Keplerian rotation curve to these data, so we only overplot the expected Keplerian curves for the selected masses and do not attempt to obtain a radius from these data points.} Most of the points less than 100 au away from the protostar fall between the 0.2 and 0.3 \Msun curves. However, the large uncertainty of the points that are closer to the protostar are too large to clearly distinguish between both cases, some also reaching values consistent with 0.1 \Msun. We decide to use a protostellar mass of 0.2 \Msun, as this is not far from the value presented in \cite{Brassfield2011B5envmass} but still looks consistent with most of the points obtained from our Gaussian fit and the 5$\sigma$ contours of the PV diagram (Fig.~\ref{fig:c18o-pv}).

\section{Comparison between \ce{H2CO} and \ce{H^{13}CO+} \label{ap:comparisonH13COp}}

We compared the emission coming from the blueshifted component from the DBSCAN analysis (Sect. \ref{sec:dbscan-clustering}) with the \ce{H^{13}CO+} emission from \cite{vantHoff2022snowline} both in the image plane and in velocity space. We first looked at the location of emission. \cite{vantHoff2022snowline} show that \ce{H^{13}CO+} emission has a ridge-like shape toward the west of B5-IRS1. \ce{H^{13}CO+} peak emission ridge coincides with the area occupied by the blueshifted component, although emission with $S/N>3$ is more extended than the blue streamer. Spatial coincidence does not mean both molecules trace the same part of the envelope, so we looked into the kinematics of  \ce{H^{13}CO+}.

\begin{figure*}[htbp]
    \centering
    \includegraphics[width=0.52\textwidth]{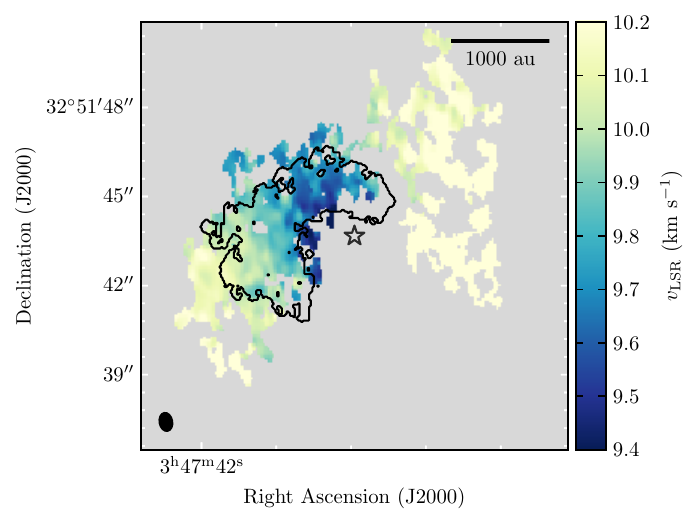}
     \includegraphics[width=0.41\textwidth]{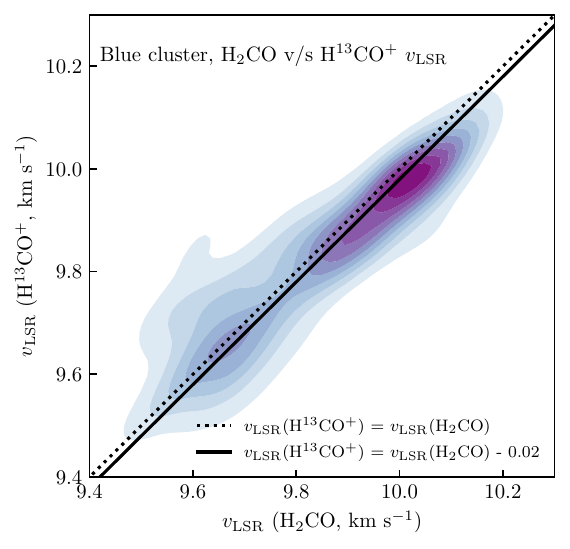}
    \caption{\refereecomment{Comparison between the \vlsr found for \ce{H^{13}CO+} observations  from \cite{vantHoff2022snowline} and our \ce{H2CO} \vlsr results. \textbf{Left:}} Central velocities obtained through the Gaussian fit to \ce{H^{13}CO+} observations from \cite{vantHoff2022snowline}. The black ellipse represents the beam size. \refereecomment{The} black contour indicates the area covered by the \refereecomment{\ce{H2CO}} blueshifted \refereecomment{cluster} found using DBSCAN. \refereetwo{The black star represents the position of B5-IRS1.} \refereecomment{\textbf{Right:}} KDE of the difference between the central velocities of \ce{H2CO} and \ce{H^{13}CO+} emission. The dotted line shows the one-to-one relation, and the solid line shows the same relation but displaced by 0.02 \kms}
    \label{fig:H13COp-vel}
\end{figure*}

 We fit a Gaussian curve to the \ce{H^{13}CO+} spectra with the same procedure as for \ce{HC3N} (Appendix \ref{ap:gaussfit}) but first selecting all spectra with $S/N>3$ to make the mask. The resulting central velocities are in Fig.~\ref{fig:H13COp-vel} \refereecomment{Left}. We note that all the velocities in the east side of B5-IRS1 are blueshifted with respect to the protostar. \ce{H^{13}CO+} shows a velocity gradient from 10.1 to 9.4 \kms as distance to the protostar decreases. This gradient is similar to the gradient seen in the blueshifted component in Fig.~\ref{fig:dbscan-redblue-velocities}, which suggest both molecules trace an infall motion. This also confirms that \ce{H^{13}CO+} does not trace outflow motion, using the same arguments as in Sect. \ref{sec:slmodel}.  
 
 We calculated the difference between \vlsr for \ce{H2CO} and \ce{H^{13}CO+} spectra to investigate if \ce{H2CO} has any additional movement with respect to \ce{H^{13}CO+} or if differences are random. For this, we reprojected the spatial grid of \ce{H2CO} to the one of \ce{H^{13}CO+} using the astropy package \texttt{reproject}. The KDE of the difference is in Fig.~\ref{fig:H13COp-vel} \refereecomment{Right}. Differences in velocity appear to be random; there appears to be no tendency for a preferential redshift of blueshift of \ce{H2CO} with respect to \ce{H^{13}CO+} for any velocity. 
 The median value of the difference is $0.02^{+0.07}_{-0.09}$ \kms, smaller than the channel width for \ce{H2CO} data (Table \ref{tab:obsprops}). This means both molecules share the same velocities and velocity gradients. We therefore conclude that both \ce{H2CO} and \ce{H^{13}CO+} trace the infall motion of the blueshifted streamer.

\end{appendix}

\end{document}